\documentclass[a4paper,twocolumn,11pt]{quantumarticle}
\pdfoutput=1
\usepackage[utf8]{inputenc}
\usepackage[english]{babel}
\usepackage[T1]{fontenc}
\usepackage{amsmath}
\usepackage{amssymb}
\usepackage{url}
\usepackage{hyperref} 

\usepackage{tikz}
\usepackage{lipsum}
\usepackage{booktabs}
\usepackage{graphicx}

\begin{document}

\title{Hybrid actor-critic algorithm for quantum reinforcement learning at CERN beam lines}

\author[1]{Michael Schenk}\email{michael.schenk@cern.ch}
\author[2]{Elías F. Combarro}\email{efernandezca@uniovi.es}
\author[1]{Michele Grossi}\email{michele.grossi@cern.ch}
\author[1]{Verena Kain}\email{verena.kain@cern.ch}
\author[1]{Kevin Shing Bruce Li}\email{kevin.shing.bruce.li@cern.ch}
\author[3]{Mircea-Marian Popa}\email{popa\_mircea95@yahoo.com}
\author[1]{Sofia Vallecorsa}\email{sofia.vallecorsa@cern.ch}

\affil[1]{European Organisation for Nuclear Research, Espl. des Particules 1, 1211 Meyrin, Switzerland}
\affil[2]{Computer Science Department - University of Oviedo, C. San Francisco 3, 33003 Oviedo, Asturias, Spain}
\affil[3]{Politehnica University of Bucharest, Splaiul Independentei 313, 060042 Bucharest, Romania}

\maketitle

\begin{abstract}
Free energy-based reinforcement learning (FERL) with clamped quantum Boltzmann machines (QBM) was shown to significantly improve the learning efficiency compared to classical Q-learning with the restriction, however, to discrete state-action space environments. In this paper, the FERL approach is extended to multi-dimensional continuous state-action space environments to open the doors for a broader range of real-world applications. First, free energy-based Q-learning is studied for discrete action spaces, but continuous state spaces and the impact of experience replay on sample efficiency is assessed. In a second step, a hybrid actor-critic scheme for continuous state-action spaces is developed based on the Deep Deterministic Policy Gradient algorithm combining a classical actor network with a QBM-based critic. The results obtained with quantum annealing, both simulated and with D-Wave quantum annealing hardware, are discussed, and the performance is compared to classical reinforcement learning methods. The environments used throughout represent existing particle accelerator beam lines at the European Organisation for Nuclear Research (CERN). Among others, the hybrid actor-critic agent is evaluated on the actual electron beam line of the Advanced Plasma Wakefield Experiment (AWAKE).
\end{abstract}

\section{Introduction}\label{sec:introduction}
The European Organisation for Nuclear Research (CERN) maintains a dense and diverse physics programme with numerous experiments requiring a broad spectrum of particle beam types to be produced and provided by the accelerator complex~\cite{physAtSPS, awake, postLIU, montbaron}. Combined with the requests for higher beam intensities and smaller beam sizes at an overall improved beam quality, this makes accelerator operation more and more challenging. The way forward is to exploit automation and improved modelling to boost machine flexibility, availability, and beam reproducibility. Ideally, as-built reversible physics models are available in the control rooms to adjust beam parameters. Whereas for many beam control problems, physics models are indeed used at the CERN accelerators, various systems are still tuned manually due to the lack of models or beam instrumentation. Recently, sample-efficient control algorithms, such as reinforcement learning (RL), have been introduced for some of these cases. Sample efficiency is essential for any optimisation algorithm in the context of accelerator operation to minimise the impact on beam time available for the physics experiments.

Q-learning is a popular RL algorithm~\cite{suttBart}, where the RL agent iteratively learns a so-called Q-function to define the optimal policy. While in classical deep Q-learning the Q-function is estimated using a deep neural network~\cite{mnih}, the free energy-based reinforcement learning (FERL) approach utilises the free energy of a coupled spin system as a Q-function approximator. The spin system is typically represented by a quantum Boltzmann machine (QBM), and its free energy is determined, for example, through quantum annealing~\cite{sallHint}. An improvement in the learning efficiency of an FERL agent compared to classical Q-learning algorithms has already been demonstrated in earlier research; however, with the restriction to discrete state-action space environments~\cite{levit, RL_QBM}.

The goals of this study are twofold. First, to remove the limitation to discrete state-action space environments and to extend FERL to multi-dimensional continuous state-action spaces, opening doors for a broader range of real-world applications. This is particularly important for particle accelerator systems where the control parameters and observables are usually defined by continuous variables. The second objective is to compare the performance of the classical RL algorithms to their quantum or hybrid counterparts in terms of both sample efficiency and the number of parameters required to model the Q-function. The objectives are achieved in a two-stage approach. FERL is first extended to continuous state-space but still discrete action-space environments, and the impact of experience replay on the sample efficiency is studied. This is done for a one-dimensional beam steering environment of an existing beam line at CERN. In a second stage, a hybrid actor-critic scheme based on the classical Deep Deterministic Policy Gradient (DDPG) RL algorithm~\cite{ddpg} is developed by combining a classical actor network with a QBM-based critic to allow for continuous state-action space environments. The hybrid scheme is validated and compared to its classical counterpart by applying it to a ten-dimensional environment of the electron beam line at the CERN AWAKE facility both in simulations and in the real world~\cite{awake}. 

The rest of the paper is organised as follows. First, the main concepts of the RL and FERL domains are introduced in Section~\ref{sec:background}, including an overview of existing related research. The contributions are discussed in Section~\ref{sec:contribution}, comprising the development of the new hybrid actor-critic algorithm and experimental results for trajectory control at the two CERN beam lines with different complexities. This includes comparisons to the performance of the corresponding classical RL algorithms both using simulated quantum annealing and D-Wave quantum annealing hardware, and an evaluation of a trained hybrid actor-critic agent on the real-world AWAKE electron beam line. Finally, some conclusions are raised, and some ideas for further study are proposed. 

\section{Background} \label{sec:background}
\subsection{Reinforcement learning}
\begin{figure}
    \centering
    \includegraphics[width=80mm]{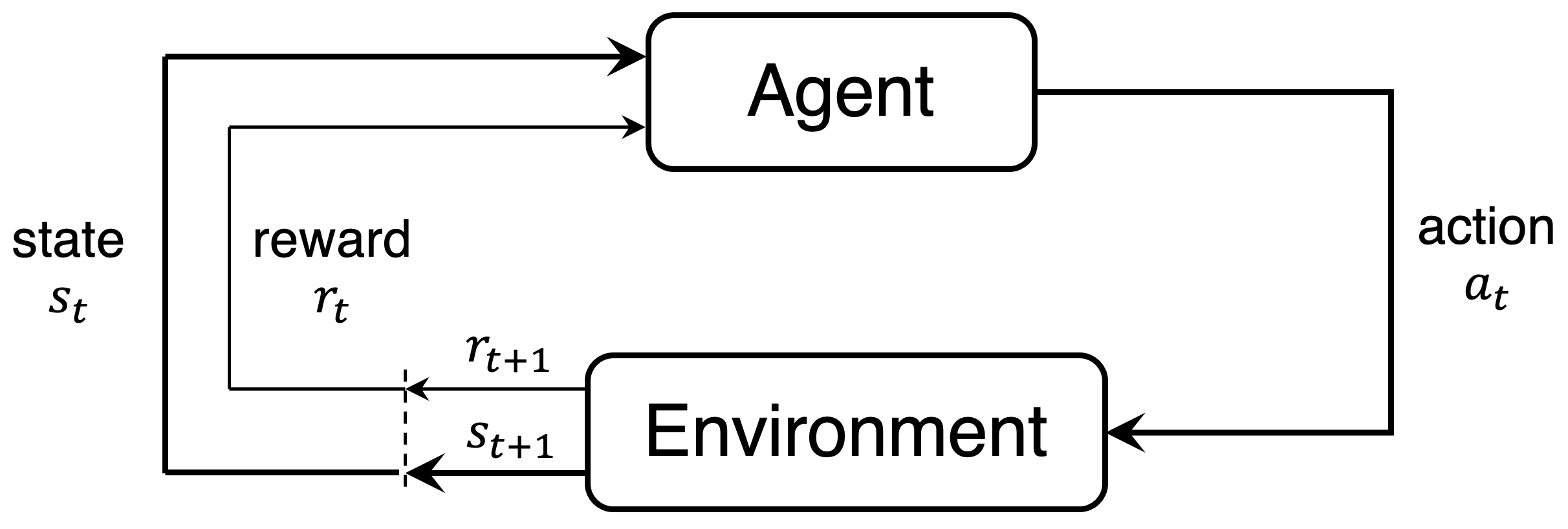}
    \caption{Reinforcement learning paradigm: the interaction between agent and environment occurs in discrete time steps. At every time step $t$, the agent observes the environment in its current state $s_t \in S$ and takes an action $a_t \in A$ following policy $\pi$. The environment transitions from state $s_t$ to $s_{t+1}$, according to the state transition probability and emits a reward $r_{t+1}$~\cite{suttBart}.}
    \label{fig:rl}
\end{figure}

\begin{figure*}
    \centering
    \includegraphics[width=130mm]{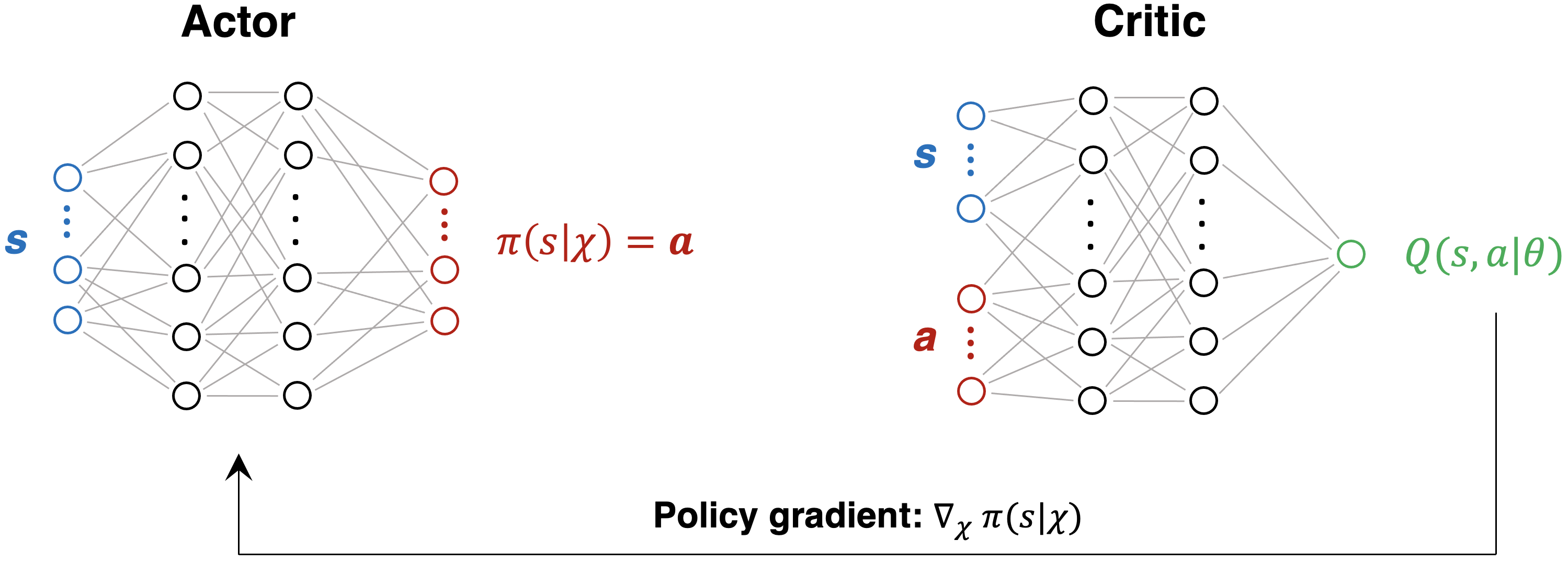}
    \caption{RL actor-critic scheme: the policy- or actor-network proposes action $a$ given a state $s$, while the critic network estimates the Q-value of the proposed state-action pair. The actor network is trained through the policy gradient rule.}
    \label{fig:actor_critic}
\end{figure*}

Reinforcement learning algorithms solve sequential decision-making problems by deciding which action $a \in A$ to take given a specific state $s \in S$~\cite{suttBart}. The problem can formally be described by using a discrete-time Markov Decision Process (MDP) consisting of state space $S$, action space $A$, state transition probabilities $P_a(s, s’)$, and immediate transition rewards $R_a(s, s’)$ emitted when moving from state $s$ to $s’$ under action $a$. The decision-making strategy is formally described through a policy function $\pi: S \times A \rightarrow [0, 1]$, which is a probability distribution that can be used to map each state to a chosen action. The optimal policy is learned through interaction with the environment to maximise the cumulative reward along the path of the visited states. The entire RL process is illustrated in Fig.~\ref{fig:rl}.

The RL algorithms discussed in this paper belong to the class of model-free RL algorithms, where the dynamics model of the environment under consideration is not explicitly learned or available. Within model-free algorithms, the class of policy optimisation methods learns the policy directly through policy gradient and is suitable for continuous action spaces. On the other hand, Q-learning methods seek to learn the so-called action-value or Q-function. The Q-function is a measure of the expected sum of discounted future rewards, assuming the agent in state $s$ takes action $a$ and then continues until the end of the episode following policy $\pi$. It is defined as
\begin{align}
    Q(s,a\vert\theta) &= \mathbb{E}_\pi \left[\sum_{k=1}^{N} \gamma^{k-1} r_{t+k}\vert s_t=s, a_t=a\right],
    \label{eq:Q_function_definition}
\end{align}
\noindent where $N$ is the number of states visited, starting from state $s=s_t$ at time step $t$ and stopping at a terminal state at the end of the episode at time step $t + N$, $\gamma \in [0, 1]$ is a discount factor, and $r_i$ is the reward the agent receives during iteration $i$. The Q-function is parameterised by $\theta$, which denotes, for example, the network weights in case the Q-function is approximated by an artificial neural network.

The Q-function can be learned iteratively by using the training samples collected during agent-environment interactions and by employing the temporal-difference rule~\cite{suttBart}
\begin{align}
    \nonumber Q(s_t, a_t)& \leftarrow Q(s_t, a_t) + \alpha [r_{t+1} \\
    &+ \gamma\, \text{max}_{a'}\, Q(s_{t+1}, a') - Q(s_t, a_t)],
    \label{eq:Q_function_update}
\end{align}
at time step $t$, where $\alpha$ is the learning rate. Once trained, i.e.\ during exploitation, a Q-learning agent always takes the action that maximises the Q-function (`greedy policy').

One of the advantages of Q-learning in comparison to policy gradient methods is its sample efficiency due to experience replay~\cite{experienceReplay}. All the state transitions, the chosen actions, and obtained rewards are stored in a replay buffer. During agent training, the Q-function updates are calculated based on mini-batches sampled from this buffer.

The Q-learning method is specifically suitable for discrete action spaces. Actor-critic algorithms combine Q-learning with policy optimisation and hence offer a relatively sample-efficient method for continuous action spaces~\cite{suttBart}.

\subsection{Deep Q-learning}
The Q-function is typically approximated by a (deep) neural network for most real-world applications where the state-action space is large. Deep Q-learning or DQN is one of the most popular Q-learning algorithms~\cite{mnih}. DQN expects the state vector as input and provides the Q-values for all possible discrete actions at the output layer. Sorting allows determining the action that maximises the Q-function efficiently.

\subsection{Actor-critic algorithms}
One major limitation of Q-learning is its restriction to discrete action-space environments. This is because the action for a given state is obtained with $a' = \arg \max_a Q(s,a)$ at inference and the update rule for agent training also needs to find the maximum over $Q(s,a')$ for all possible actions $a'$ (see Eq.~\ref{eq:Q_function_update}).

The actor-critic scheme removes this limitation. The most basic algorithm is Deep Deterministic Policy Gradient (DDPG)~\cite{ddpg} with its architecture illustrated in Fig.~\ref{fig:actor_critic}. It consists of a ``critic'' network parameterised by weights $\theta$ to approximate the Q-function and an ``actor'' network parameterised by weights $\chi$ to learn the policy. According to this scheme, the actor proposes a (continuous) action given the current state, and the critic provides feedback on how good that state-action pair is by calculating its Q-value. DDPG interleaves Q-learning and policy updates. The update rule for the critic network is identical to Eq.~\ref{eq:Q_function_update} except that the term $\text{max}_{a'}\, Q(s_{t+1}, a')$ is now replaced by $ Q(s_{t+1}, \pi (s_{t+1}))$. Following the chain rule, the policy gradient updates can be calculated as
\begin{align}
    \nonumber \nabla_{\chi} \pi &= \mathbb{E}_\pi \left[\nabla_{\chi} Q(s, \pi(s\vert\chi)\vert \theta)\right] \\ &= \mathbb{E}_\pi \left[ \nabla_a Q(s, a \vert \theta) \cdot \nabla_{\chi} \pi(s\vert \chi)\right]. \label{eq:policy_gradient}
\end{align}

\subsection{Energy-based reinforcement learning}
One can employ energy-based models to approximate and learn the Q-value function as an alternative to using a standard feed-forward neural network. In an energy-based model, the Q-function is defined through the energy function of a statistical system. Every configuration of that system has a specific energy associated with it that can be used as a Q-value estimate. By tuning the system's parameters, an energy function is adjusted iteratively to best approximate the Q-function of the RL task under consideration. A particularly well-suited example of a statistical system for FERL is the Boltzmann machine, an energy-based model represented by a probabilistic network of binary variables~\cite{boltzmannMachine}.

\subsubsection{Clamped quantum Boltzmann machine} \label{sec:clamped_QBM}
A Boltzmann machine (BM) consists of visible and hidden (latent) variables. They are denoted by $v_i$, with $i \in V$,  and by $h_j$, with $j \in H$, respectively. Visible variables typically serve as the inputs and outputs, and hidden variables are added to increase the expressivity of the model. 

For the FERL algorithms discussed in this paper, the BM's topology is chosen to be clamped, building upon the research in~\cite{levit}. This means that the visible variables are not part of the network. Instead, they enter the system's energy function as biases or self-couplings to the specific hidden variables assigned to them. These bias terms are used in a weighted sum, where the weights are given by the coupling strengths between visible and hidden variables. Bias terms are linear in the state of the hidden variable with which they are associated. On the other hand, couplings between hidden variables, such as $w_{jk} \in \mathbb{R}$, $j, k \in H$, correspond to quadratic contributors to the energy function.

\begin{figure}
    \centering
    \includegraphics[width=75mm]{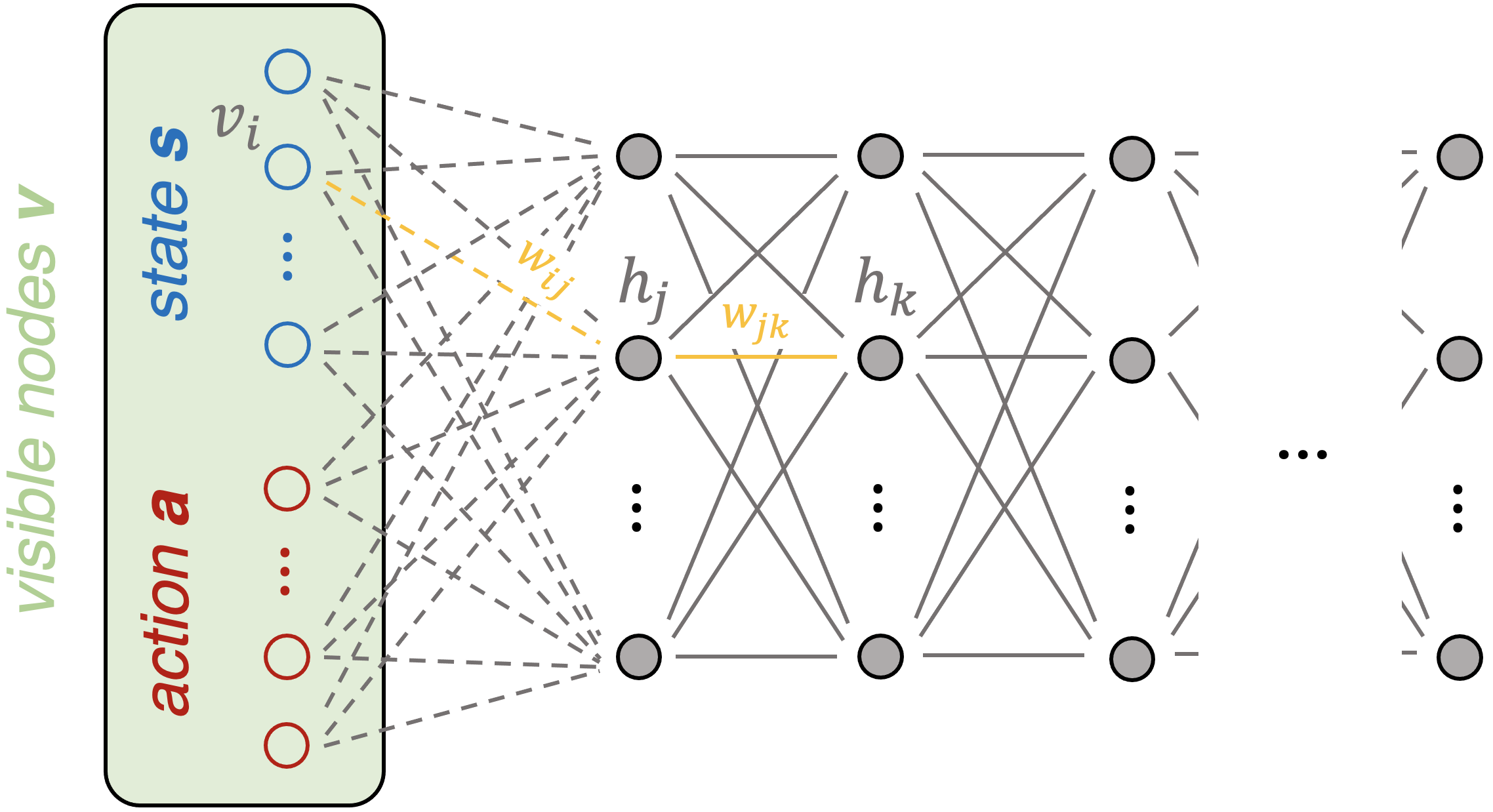}
    \caption{Clamped BM with topology typically used for FERL applications. Visible nodes are given by the state-action pair (green box). Hidden nodes are shown in grey. Visible-hidden and hidden-hidden couplings are illustrated by dashed and solid lines, respectively.}
    \label{fig:clamped_qbm}
\end{figure}

For RL, the visible variables are composed of the vectors of the state-action pair $\mathbf{v} = (s, a) \in \mathbb{R}^{d_S + d_A}$, where $d_S$ and $d_A$ are the dimensionalities of the state and action space, respectively. Figure~\ref{fig:clamped_qbm} illustrates a clamped BM typically employed for FERL.

A quantum BM (QBM) can be represented by a physical system of coupled qubits in the presence of a purely transverse magnetic field, here pointing along the $x$-axis. Each node of the BM can assume a spin ``up'' or ``down'' state following a certain probability distribution. The system's energy states are described by the Hamiltonian of the transverse-field Ising model~\cite{transverseIsingModel}
\begin{align}
    \nonumber \mathcal{H}(\mathbf{v}) = &-\sum_{\substack{i\in V,\\ j \in H}} w_{ij} v_i \sigma^z_{h_j} - \sum_{j, k \in H} w_{jk} \sigma^z_{h_j} \sigma^z_{h_k} \\
    &- \Gamma \sum_{j \in H} \sigma^x_{h_j},
    \label{eq:ising}
\end{align}
where $\sigma^{x}_{h_j}$ and $\sigma^{z}_{h_j}$ are the Pauli spin matrices acting on the hidden node $h_j$ for the $x$- and $z$-directions, respectively, and $\Gamma$ denotes the transverse magnetic field strength. The transverse field introduces quantum fluctuations which enable, among others, quantum tunnelling~\cite{QAspins}.

When measuring the spin states along the $z$-coordinate, one loses information about the spin $x$ components. This can be overcome by replica stacking, a method developed in~\cite{levit} to expand the transverse-field Ising model based on the Suzuki-Trotter expansion. By applying this technique, the non-zero transverse-field Ising model can be represented by a classical Ising model of one dimension higher with an effective Hamiltonian given by
\begin{align}
    \nonumber \mathcal{H}^\text{eff}&(\mathbf{v}) = \\
    \nonumber &-\frac{1}{N_r} \sum_{l=1}^{N_r} \left(\sum_{j, k \in H} w_{jk} h_{j,l} h_{k,l}  + \sum_{\substack{i\in V,\\ j\in H}} w_{ij} v_i h_{j,l}\right) \\
    &- w^+ \left(\sum_{j\in H} \sum_{l=1}^{N_r} h_{j,l} h_{j,l+1} + \sum_{j \in H} h_{j,1} h_{j,N_r} \right),
\end{align}
with $w^+ = \frac{1}{2\beta} \log \left[\coth \frac{\Gamma \beta}{N_r}\right]$, where $N_r$ is the number of replicas, $\beta$ the inverse temperature, and $h_{j,l}$ denotes the hidden node with index $j$ in replica $l$.

For FERL, the negative free energy $F(\mathbf{v})$ of the clamped QBM is used to approximate the Q-function
\begin{align}
    \label{eq:Qfunc}
    \nonumber Q(\mathbf{v}) &\approx -F(\mathbf{v}) \\
    &= -\langle \mathcal{H}^\text{eff}(\mathbf{v})\rangle - \frac{1}{\beta} \sum_c \mathbb{P}(c\vert\mathbf{v})\, \log \mathbb{P}(c\vert\mathbf{v}),
\end{align}
where $c$ ranges over the spin configurations and $\mathbb{P}(c\vert\mathbf{v})$ denotes the probability of observing spin configuration $c$ given visible nodes $\mathbf{v}$.

The temporal difference update rules for the QBM weights follow from the Bellman equations
\begin{align}
    \label{eq:weight_updates}
    \nonumber w_{ij} &\leftarrow w_{ij} + \alpha\, \Delta Q\, v_i\, \langle h_j \rangle, \\
    w_{jk} &\leftarrow w_{jk} + \alpha\, \Delta Q\,\langle h_j h_k \rangle, 
\end{align}
with $\Delta Q = [r_{t+1} + \gamma Q(s_{t+1}, a_{t+1}) - Q(s_t, a_t)]$, and $i \in V$, $j, k \in H$, and where $\langle h_j \rangle$ and $\langle h_j h_k \rangle$ are the expectation values of the hidden nodes and of the products of hidden nodes, respectively. Based on Eq.~\ref{eq:weight_updates}, the weights of the QBM can be learned iteratively in analogy to classical Q-learning.

\subsubsection{Quantum annealing}
Quantum adiabatic computing~\cite{farhi2000quantum} is a model of quantum computation in which the adiabatic theorem~\cite{born1928beweis} is exploited to obtain the ground state of a given Hamiltonian $H_f$. Initially, the ground state of a simple Hamiltonian $H_i$ is prepared. Then, the system is evolved according to the time-dependent Hamiltonian
$$H(t) = A(t)H_i + B(t)H_f,$$
where $A(t), B(t): [0,T] \rightarrow \mathbb{R}$ are such that $A(0)=B(T)=1$ and $A(T)=B(0)=0$. Provided that the evolution is slow enough, the adiabatic theorem ensures that, starting from the ground state of $H(0)=H_i$, the system will end up in the ground state of $H(T)=H_f$.   

In practice, however, the time $T$ can be extremely large (and, in some cases, very expensive or even impossible to compute~\cite{cubitt2015undecidability}). For this reason, it is common to apply a heuristic implementation of quantum adiabatic computing that goes by the name of quantum annealing~\cite{geoch}. In quantum annealing, $H_f$ is taken from a restricted family of Hamiltonians, usually those of the transverse-field Ising model, and $T$ is fixed to some constant time, even if it does not guarantee adiabaticity.

Quantum annealing is implemented by the Canadian company D-Wave in a series of quantum devices, some of which can be accessed online through a cloud-based service~\cite{leap}. One of their most popular uses is the approximation of solutions to combinatorial optimisation problems. This is achieved by mapping the objective function to a Hamiltonian whose ground states are minimum-cost solutions to the original problem (see~\cite{geoch} for details). In this work, following~\cite{levit}, quantum annealing will be used to estimate the free energy of quantum Boltzmann machines which serve as Q-function approximators, as explained in the previous subsection.

\subsection{Related work}
\label{sec:related}
Quantum machine learning (QML)~\cite{biamonte-qml} is a discipline aiming to establish a productive interplay between the parallel revolutions of quantum computing and artificial intelligence. Among various machine learning tasks and methods widely described in the literature, the quantum computing connection to the field of RL is still in a preliminary state, whereas actor-critic methods are among the state-of-the-art in current classical RL literature.
With no ambition for completeness, but with a cross-sectional view of the literature, in this section, previous works in the field that made steps towards connecting quantum computing and RL are summarised. 

Quantum enhancements for RL can be found mainly in sample complexity and the acceleration of the algorithmic part of the learning agent. However, the importance of probing speed-up has only been studied for special models~\cite{RL_QBM,PhysRevX.4.031002} and by building on quantum annealing systems like those developed by D-Wave~\cite{dwavedocs}.

Additional consideration should be made concerning the technology adopted: different approaches are linked to different quantum formulations of the assigned problem, either quantum annealing or gate-based quantum computation.
Some proposed algorithms can show a proved speed-up only in the fault-tolerant quantum regime. Current near-term quantum devices cannot handle complex quantum routines, so alternatives like the combination of deep BMs and quantum BMs or parameterised quantum circuits (PQC) for variational quantum algorithms (VQA) are being investigated for heuristic quantum speed-ups. Considering the commonly accepted division of RL algorithms into policy-based and value-based methods, a short summary of PQC implementation is provided in the following.

A first step in assessing PQCs in policy-based RL algorithms has been made in \cite{jerbi}, where the authors focus on the role of data-encoding and readout strategies. Regarding the value-based approach, there are fundamental questions about the applicability of VQAs, particularly for Q-learning algorithms, where the role of the deep Q-network is to serve as a function approximator of the Q-function.
Again, in \cite{jerbi} for the Q-learning approach with policy gradient-based RL, the authors identify families of environments that are proven to be hard for any classical learner but can be solved in polynomial time by a quantum learner in a policy learning setting.

In \cite{VQCDRL16}, a two discrete-state environments solution with PQC is studied, where a layer of additive weights follows the quantum model's output.
Instead, in \cite{RLQVC17}, the authors propose a continuous and discrete state spaces implementation where angle encoding is used for the continuous part with one initial layer of rotation gates. 
Nevertheless, none of the continuous state-space environments run for the Cart Pole benchmark achieve a satisfactory convergence according to its original specification.

Another important reference that provides an extended discussion about the state of the art for QRL is~\cite{Jerbi_2021}. From the theoretical point of view, the authors explore the possibility of approximating functions and the complexity of sampling from different models. 
This led to quantum generalisations of classical energy-based models. They are quantum enhancements for a class of deep RL which can have further advantages over conventional methods by their capacity to capture more complex action-state distributions.
Beyond general considerations, also for this case, the authors consider mainly one (simple) classical RL architecture, i.e., deep Q-learning, which achieves better learning performances than conventional methods when the state and action spaces are large.

As far as annealing is concerned, concerning enhancements of the algorithmic part of the learning agent, speed-ups have been explored by building on the system of D-Wave computers. In \cite{levit} the authors implemented free energy-based reinforcement learning on the D-Wave 2000Q machine, where they proved that replica stacking is a successful method for estimating effective classical configurations obtained from a quantum annealer, with significant improvement over FERL using classical Boltzmann machines. 
In \cite{dwave66}, policy evaluation for discrete state spaces is embedded in the D-Wave QPU (Quantum Processing Unit). The authors also look into dealing with continuous state space in the context of self-driving vehicles. However, despite the interesting proposal, the work does not guarantee speed-up.

To the best of our knowledge, regardless of the computing paradigm considered, no previous proposal of a hybrid (i.e.\ quantum-classical) actor-critic scheme is present in the literature.

\section{Contribution} \label{sec:contribution}
\subsection{Objectives and study cases}
The main contribution of this paper is to develop and study a hybrid actor-critic (A-C) scheme, combining a classical policy network with a quantum-based Q-network represented by a QBM. Given the improved sample efficiency observed empirically for FERL Q-learning on discrete state-action tasks~\cite{levit, RL_QBM}, the aim is to study whether an improvement in the training efficiency can also be observed for a hybrid A-C algorithm. The work was initially motivated by the existing control problems at the CERN accelerator complex, which would greatly benefit from sample-efficient RL algorithms and whose control variables and observations are usually defined over continuous spaces.

The environments under study are two existing CERN beam lines representing control problems of different degrees of complexity. The first describes a target steering task for a proton beam with a single control variable. In contrast, the second one represents the AWAKE electron beam line featuring ten control parameters. For both tasks, accurate simulations exist, which will be used mostly to train and study the different algorithms. The environments are built on top of the \textsc{OpenAI gym}~\cite{openai} template to facilitate the use of existing classical RL algorithm implementations, such as \textsc{Stable-Baselines3}~\cite{stable-baselines3}.

In the following, two studies are presented. Study A considers an FERL Q-learning agent to solve a one-dimensional target steering problem. The impact of experience replay on the sample efficiency is assessed and compared to classical deep Q-learning. The action space is discretised, but the state space is continuous. In Study B, the performance of the hybrid A-C is evaluated for the ten-dimensional electron beam line and is compared to the classical DDPG algorithm using a continuous state-action space.

All the studies are first performed using the simulated quantum annealing (SQA) library \textsc{Sqaod}~\cite{sqaod}. Eventually, however, the FERL Q-learning and hybrid A-C agents are also adapted to and trained on D-Wave Advantage quantum annealing hardware (QA)~\cite{dwavedocs}. Furthermore, for the AWAKE beam line environment, evaluations of the RL agents can also be performed in the real world, as equipment safety is less of a concern given the type and parameters of the particle beam. Hence, the agents trained in the simulated environment are also deployed on the real AWAKE beam line (sim-to-real transfer).

\subsection{Hyperparameter selection} \label{sec:procedure}
To guarantee a fair comparison between the classical RL and quantum FERL approaches, extensive hyperparameter searches were performed to ensure that the agents always train with the best sample efficiency possible. The search was realised by means of the \textsc{Ray Tune}~\cite{raytune} and \textsc{Optuna}~\cite{optuna} libraries. $500$ hyperparameter sets were explored for every case. For every parameter set, the RL agents were trained $15$ times from scratch, and their performance was evaluated based on $500$ randomly initialised episodes. The following hyperparameters were included in the search: 
\begin{itemize}
    \item \textbf{Exploration parameters:} initial $\varepsilon$-greedy and the fraction of training period with $\varepsilon$-greedy policy, maximum allowed iterations per episode;
    \item \textbf{Learning parameters:} initial learning rate, batch size, reward discount factor, target-net soft update factor;
    \item \textbf{Annealing parameters:} inverse annealing temperature, final transverse field.
\end{itemize}

The annealing time was fixed at $100$ steps (SQA), and the number of Trotter slices was set to $5$. The same procedure was employed for all the studies discussed in the following.

\subsection{Study A: FERL Q-learning with continuous state space}
This section discusses the performance of the classical and FERL agents on a one-dimensional proton beam target steering environment. The FERL method described in~\cite{levit}, where the authors work with discrete binary state-action space environments, is extended to continuous state space. This is possible in a clamped QBM where the visible state input nodes only enter as bias terms.

The sample efficiencies of FERL and classical Q-learning agents and the impact of experience replay on the latter are assessed. Furthermore, the sample efficiency is also measured depending on the complexity of the Q-network. Initially, this task was also formulated as a discrete, binary state-action space environment, reproducing the method employed in~\cite{levit}. For completeness, those results are included in Appendix~\ref{app:1ddiscretebinary}.

\subsubsection{Environment and RL task} \label{sec:1dsteering}
\begin{figure*}
    \centering
    \includegraphics[width=145mm]{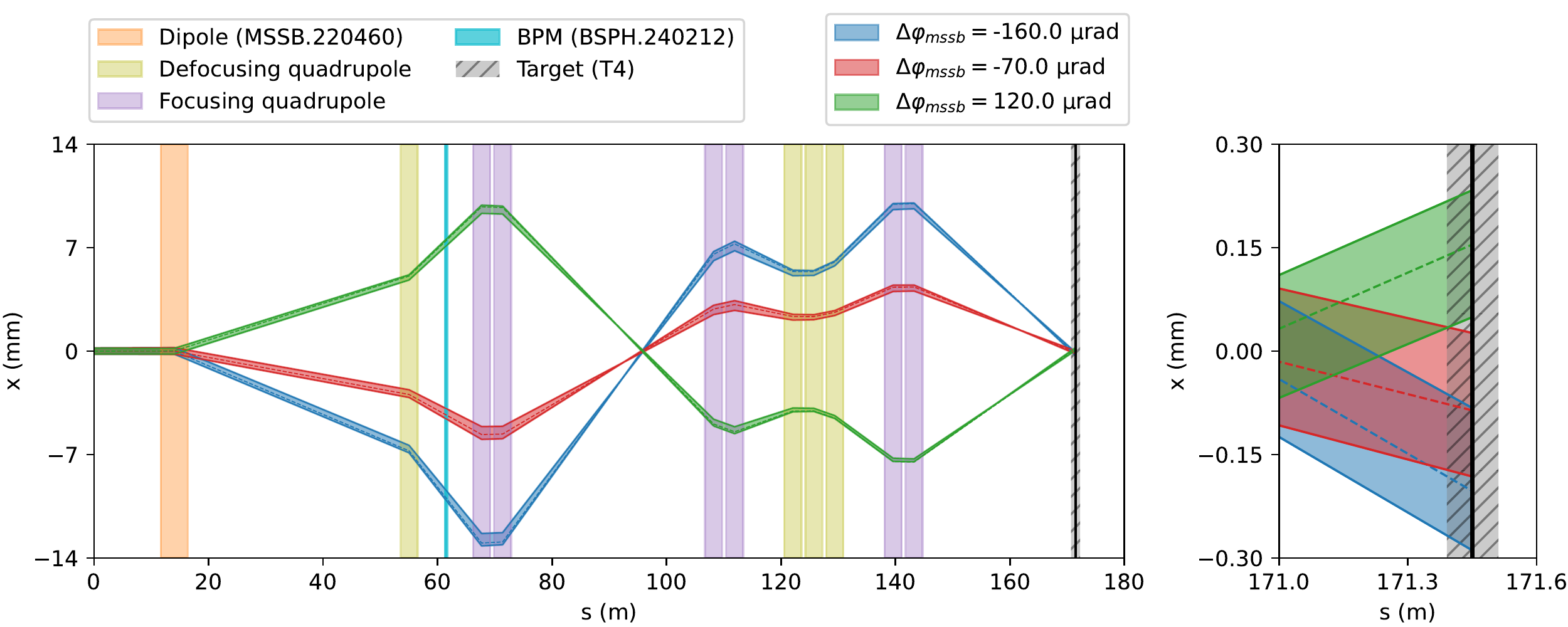}
    \caption{One-dimensional beam target steering task at the CERN TT24-T4 beam line. \textit{Left:} Horizontal beam trajectories obtained from tracking simulations are shown for three different settings of the main deflecting dipole (orange). \textit{Right:} Zoomed view on the target (grey, hatched) region showing the horizontal position of impact of the beam for the three settings of the main dipole.}
    \label{fig:1d_env}
\end{figure*}

The one-dimensional beam target steering environment is based on the beam optics of the TT24-T4 transfer line at CERN~\cite{tt24t4}. This line is about $170\,\text{m}$ long and transports protons with a momentum of $400\,\text{GeV/c}$ from the Super Proton Synchrotron (SPS) to some of the fixed-target physics experiments installed in the CERN North Area. TT24 is equipped with several dipole and quadrupole magnets to steer and focus the beam, various beam position monitors (BPM), and the actual target, which is placed at the end of the line. The objective of the task is to optimise the number of particles hitting the target by tuning the first dipole magnet in the line to maximise the event rates in the particle detectors.

The left-hand side of Fig.~\ref{fig:1d_env} shows the relevant elements of TT24 together with horizontal beam trajectories obtained from tracking simulations for three different settings of the main bending dipole (orange). Depending on the dipole deflection angle, the particles hit the target (grey, hatched) at different horizontal positions, as illustrated by the zoomed view on the right-hand side of the figure. There are focusing (purple) and defocusing (olive) quadrupoles along the beam line to keep the beam particles confined.

The RL task is formalised as follows. The state $s$ is defined by the beam position reading of one of the BPMs installed in the beam line (cyan). The action is given by the relative deflection angle induced by the dipole magnet (orange). Two discrete actions are possible -- either increasing or decreasing the deflection angle by a small, fixed amount of $15\,$\textmu rad. The allowed range of deflection angles is $[-140, 140]\,$\textmu rad. The reward is calculated as the overlap integral between the target and the beam in the interval $x \in [-3\sigma_\text{beam}, 3 \sigma_\text{beam}]$, assuming a Gaussian particle distribution with transverse rms size $\sigma_\text{beam}$.

\subsubsection{Results: classical deep Q-learning vs FERL}
\begin{figure*}
    \centering
    \includegraphics[width=65mm]{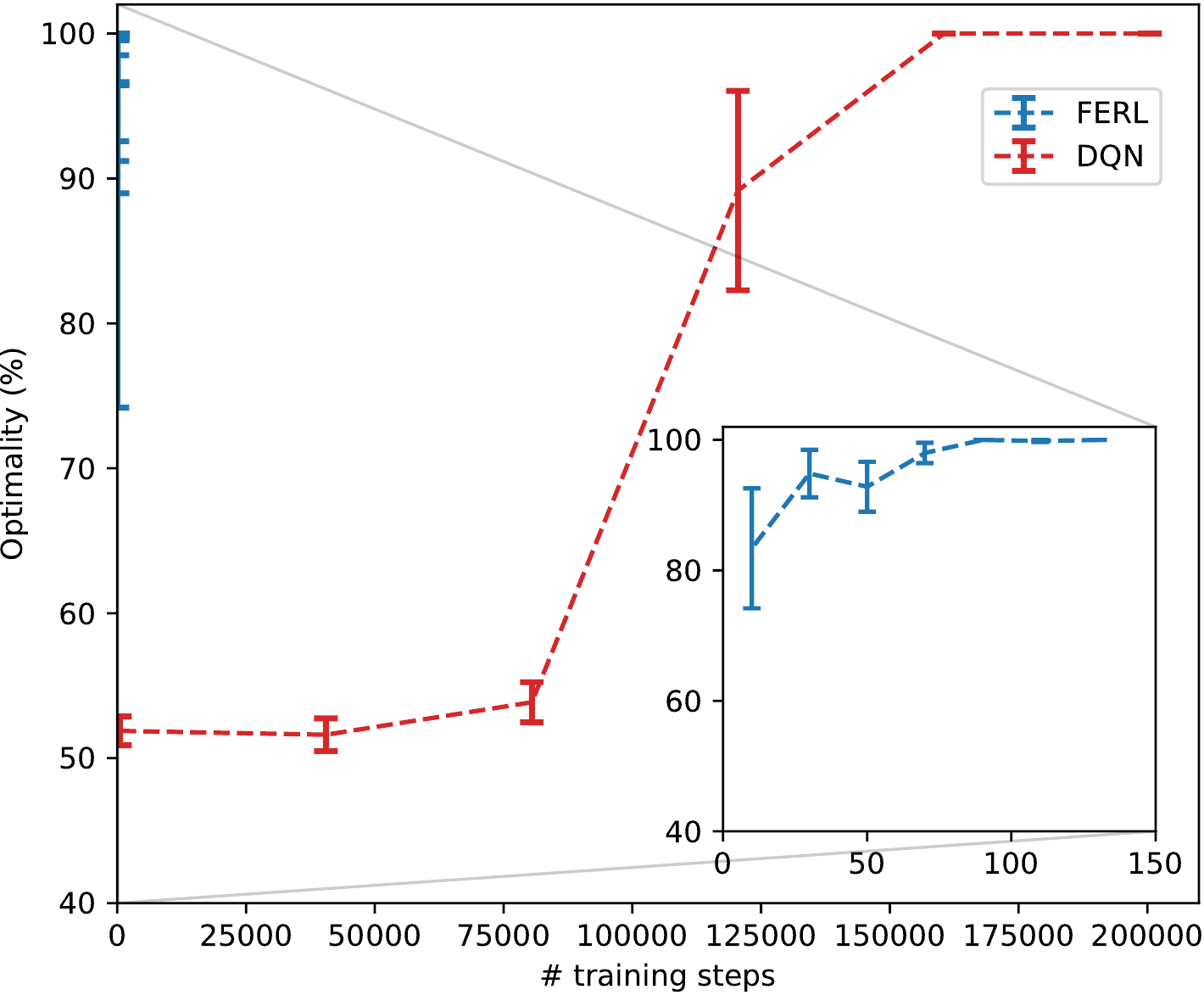}
    \hspace{0.8cm}
    \includegraphics[width=65mm]{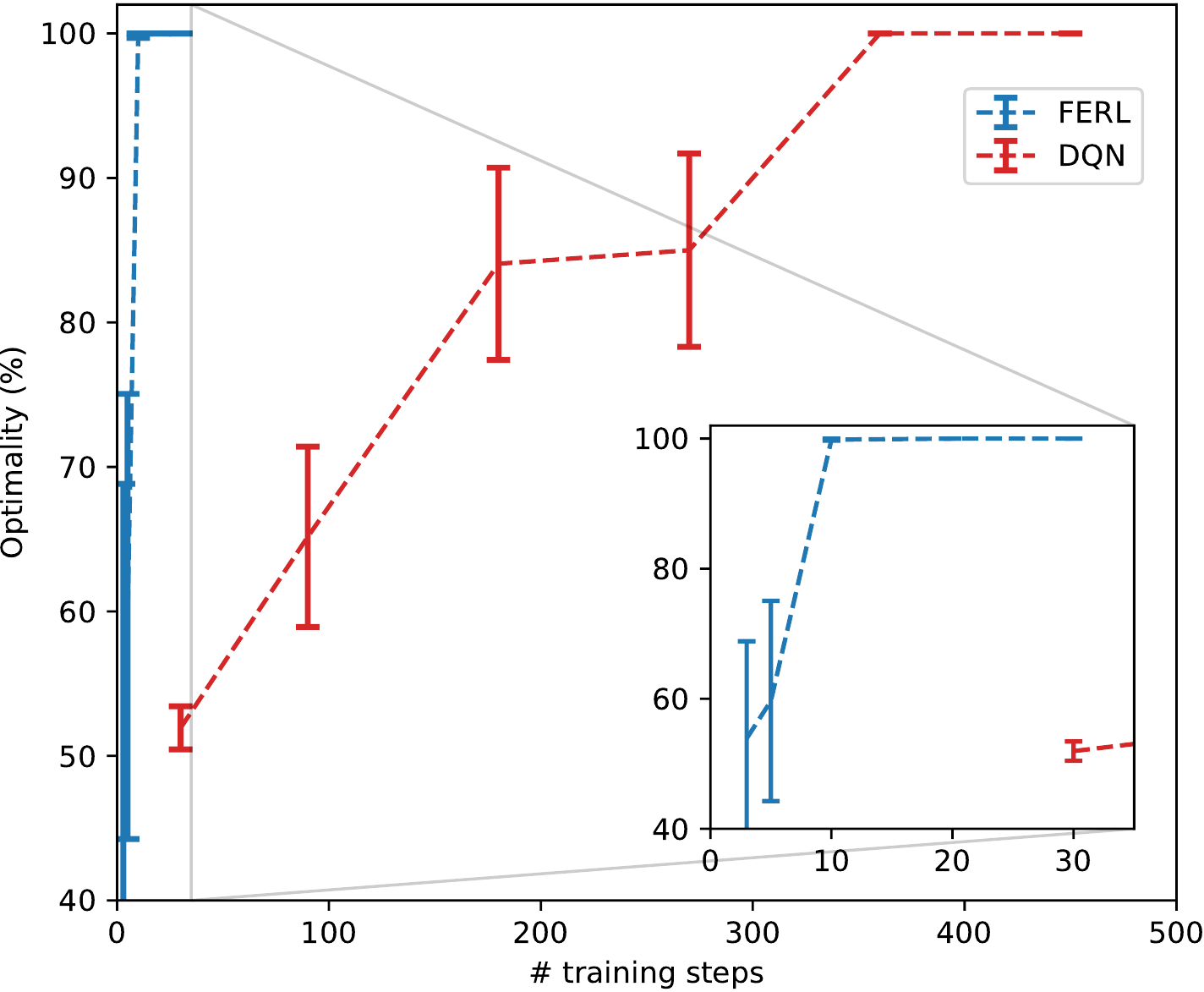}
    \caption{One-dimensional beam target steering environment with a continuous state and discrete action space: convergence study for agent optimality vs the number of training steps \emph{without} (left) and \emph{with} (right) experience replay for a classical DQN (red) and an FERL (blue) agent shown on a log-log scale.}
    \label{fig:1d_cont}
\end{figure*}

Figure~\ref{fig:1d_cont} shows convergence studies of policy optimality vs number of training steps for the classical DQN (red) and the FERL (blue) agents after performing extensive hyperparameter optimisation. The optimality metric is defined as the fraction of the state space from where the trained agent will take optimal actions to reach the reward objective with the smallest number of steps possible. The classical agent uses a DQN architecture with two hidden layers with $128$ nodes each. The number of hidden layers and the number of nodes per layer have been treated as hyperparameters, in addition to the parameters listed in Section~\ref{sec:procedure}. The same number of nodes was assumed in every layer. The FERL agent, on the other hand, uses a $1\times 2$ unit-cell Chimera graph QBM with a total of $16$ qubits following the topology of the D-Wave 2000Q quantum annealer~\cite{dwavedocs}.

\begin{figure}
    \centering
    \includegraphics[width=80mm]{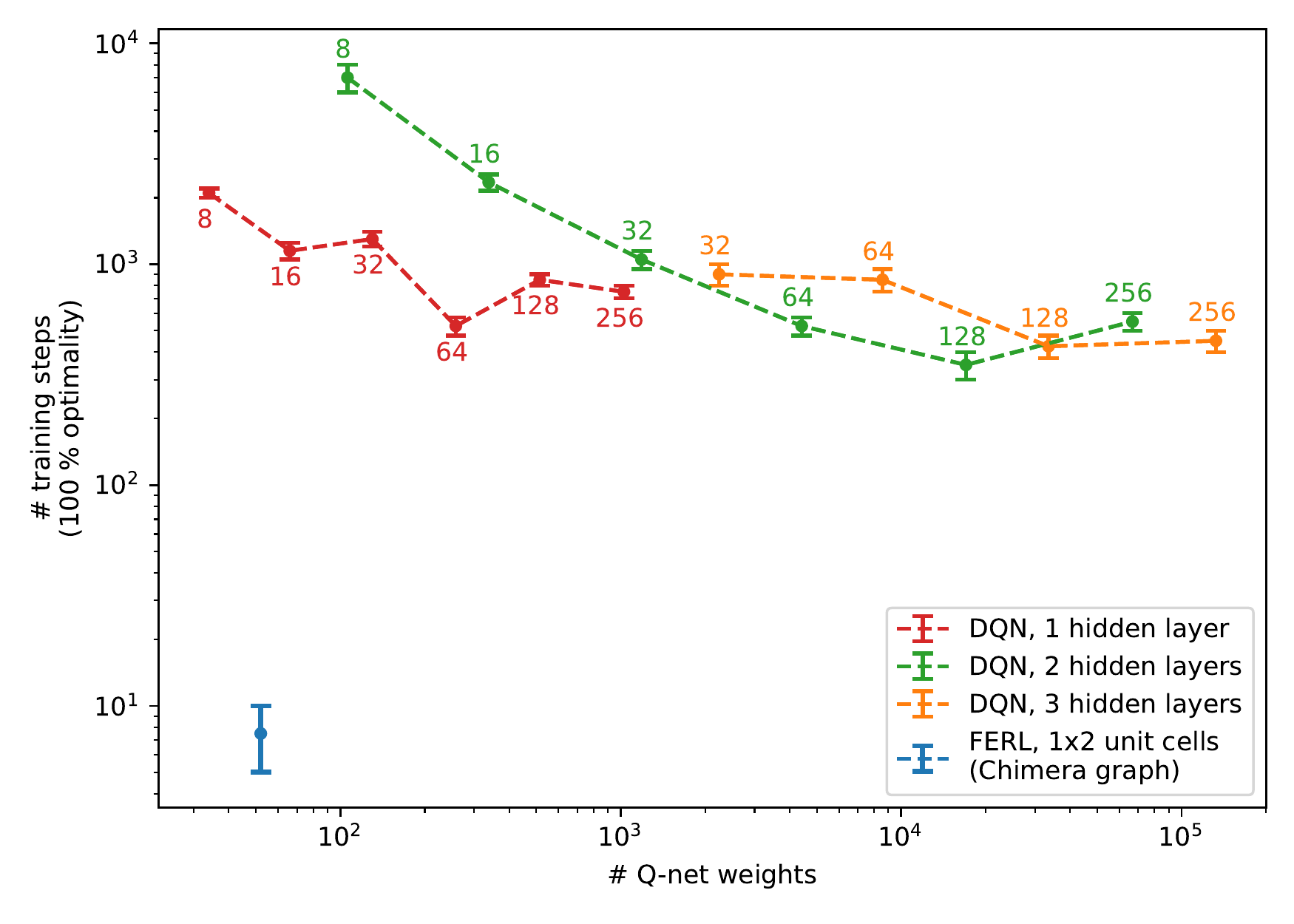}
    \caption{Required number of training steps to train a policy with $100\,\%$ optimality vs number of DQN weights for different architectures (red, green, orange) and the FERL (blue) agent, respectively. The labels next to each data point denote the number of nodes per hidden layer.}
    \label{fig:1d_descr_power}
\end{figure}

Two main conclusions can be drawn from this result. First, FERL significantly improves sample efficiency compared to the classical approach. With experience replay enabled, the number of training steps required to reliably achieve policy optimality of $100\,\%$ is $10$ steps for FERL and $380$ for DQN. Second, for the given RL task, the impact of experience replay is significant not only for classical deep Q-learning (a well-known result~\cite{experienceReplay}) but also for the FERL algorithm. Improvements in sample efficiency of a factor of $400$ for DQN and $10$ for FERL were observed when enabling experience replay. 

In addition to improved sample efficiency, the QBM employed for FERL also exhibits a higher descriptive power than the classical DQN. This is illustrated by the plot in Fig.~\ref{fig:1d_descr_power}. It shows the required number of training steps to achieve $100\,\%$ policy optimality vs the number of parameters in the Q-network on a log-log scale, with experience replay enabled. For the classical case, results for DQNs with one (red), two (green), and three (orange) layers are shown. Each data point has been obtained from a convergence scan with several training steps after individual hyperparameter optimisation. The labels next to the data points refer to the number of nodes per hidden layer. The performance of the FERL agent using a $1\times 2$ unit-cell Chimera graph described above is shown in blue. It trains an optimal policy in only $10$ steps using a QBM with $52$ weights, while the most sample-efficient classical agent trains in about $380$ steps requiring a DQN with two hidden layers and a total of about $10^4$ weights. A classical agent with just one hidden layer and $64\,\text{nodes}$ ($258$ DQN weights total) can also be trained successfully. However, it requires more than $500$ training steps to become optimal. Adding more than two hidden layers does not seem to improve the sample efficiency further for the RL task under consideration.

\subsubsection{Results: SQA vs D-Wave Advantage QA}
\begin{figure*}[t]
    \centering
    \includegraphics[width=145mm]{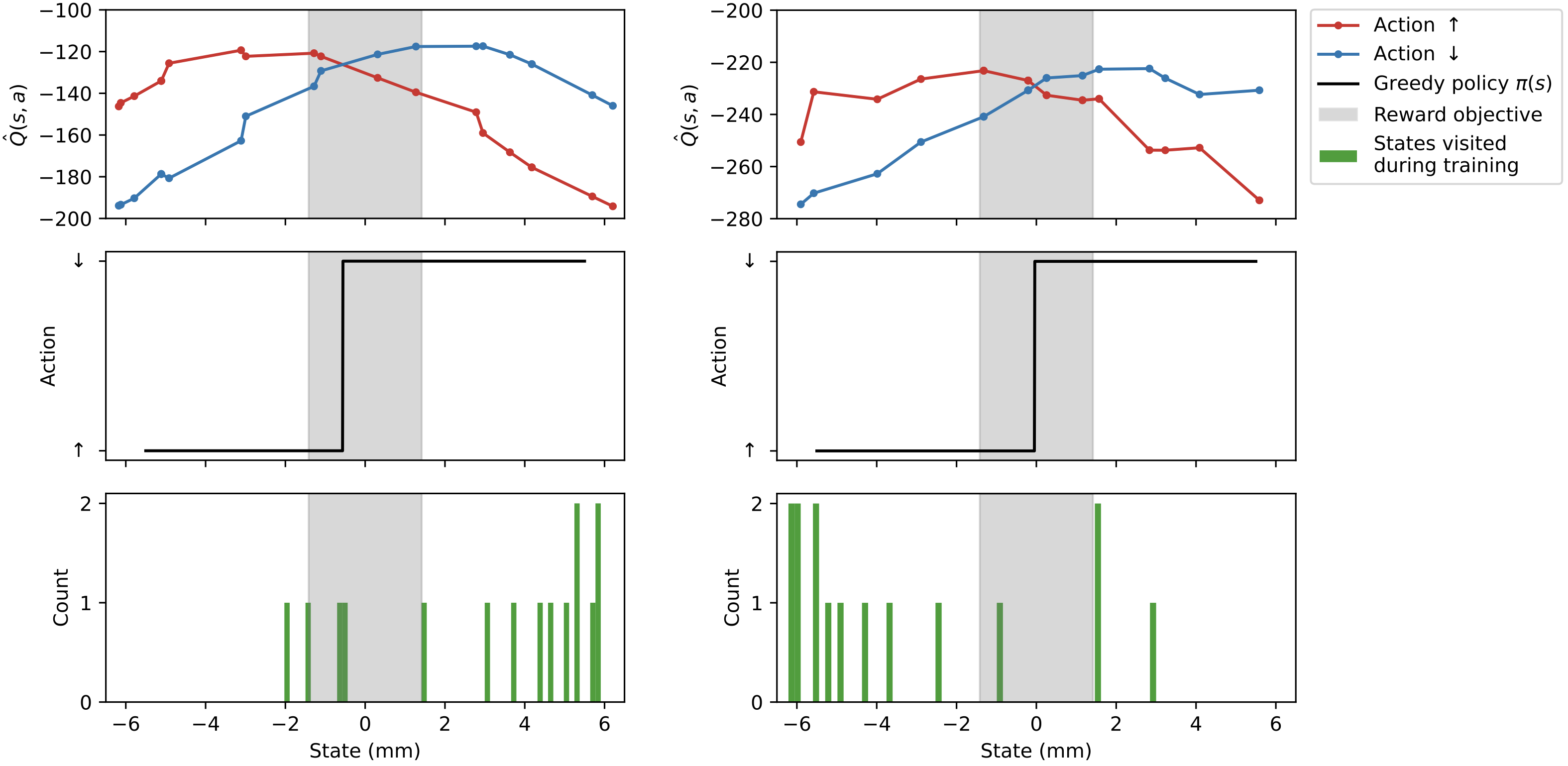}
    \caption{Q-value estimates after training (top), derived greedy policies (middle), and histograms of visited states during training (bottom) on the one-dimensional beam steering environment with continuous state space. The training was performed using SQA (left) and running on the D-Wave QA (right).}
    \label{fig:1d_cont_policy_dwave}
\end{figure*}

Figure~\ref{fig:1d_cont_policy_dwave} shows the state-action value estimates $\hat{Q}(s, a)$ (top), the derived greedy policy (middle), and histograms of the states visited during the training phase (bottom) for FERL agents trained with SQA (left) and on the D-Wave QA (right), respectively. The grey area marks the region where the beam intensity on target is beyond the set threshold, i.e.\ the reward objective is reached. 

The agent has only visited $15$ states during training in both cases. While the specific values of the learned Q-functions are not the same, in both cases, one obtains greedy policies which are $100\,\%$ optimal. The hyperparameters were tuned with SQA and were directly used on the D-Wave QA. The results demonstrate that the RL agent can be trained successfully on a real QPU. Furthermore, the number of states visited during the entire training phase is comparable to what was required using SQA (Fig.~\ref{fig:1d_cont}, right, blue curve).

\subsection{Study B: hybrid A-C scheme}
\subsubsection{Motivation and description of the algorithm}
In the A-C scheme depicted in Fig.~\ref{fig:actor_critic}, it is possible to replace the classical Q-network with a QBM. This allows combining FERL with the policy update of a classical actor network and means that only one state-action pair evaluation is needed at every iteration. 

Given the impressive results for Q-learning with FERL it is expected that the FERL critic also trains more efficiently in the hybrid A-C scheme, which should speed up the convergence of the actor network.

One main ingredient to the A-C scheme is the policy gradient evaluation (Eq.~\ref{eq:policy_gradient}) which includes the calculation of the derivative $\nabla_a Q(s,a\vert\theta)$, where $Q(s,a\vert\theta)$ is given by Eq.~\ref{eq:Qfunc}. Using numerical differentiation with finite differences is the most straightforward method of estimating this derivative without making additional assumptions on the probability distributions of spin configurations $c$. Alternatively, one could consider a semi-analytic approach. The calculation of $\partial_{\mathbf{v}=(s,a)}\, \mathcal{H}_v^\text{eff}$ is straightforward. However, the derivative of the entropy term cannot be easily evaluated except if $\mathbb{P}(c\vert\mathbf{v}) \approx 1$ for a specific spin configuration $c = c_0$, given a configuration of visible nodes $\mathbf{v}$, in which case the entropy term would be negligible. This paper does not develop the latter approach further, and numerical differentiation with finite differences is used instead.

The hybrid scheme also has the advantage that once the agent is trained, the (quantum) critic is no longer required at inference time. Since the policy is represented by a purely classical network, deploying the trained agent in a real-world environment becomes straightforward. This is particularly true in a sim-to-real RL setting, as discussed for the AWAKE beam line below.

\subsection{The CERN AWAKE facility}
The Advanced Wakefield Experiment (AWAKE) at CERN uses high-intensity $400\,\text{GeV}$ proton bunches from the Super Proton Synchrotron (SPS) as a plasma wakefield driver. Electron bunches are simultaneously steered into the plasma cell to be accelerated by the proton-induced wakefields. Electron energies up to $2\,\text{GeV}$ have been demonstrated over a plasma cell of $10\,\text{m}$ length corresponding to an electric field gradient of $200\,\text{MV/m}$~\cite{awake}. The ultimate goal for AWAKE is to reach a field gradient of $1\,\text{GV/m}$. These numbers are to be compared to conventional accelerating structures using radio-frequency (rf) cavities in the X-band regime, which are currently limited to accelerating field gradients of about $150\,\text{MV/m}$~\cite{xband}.

\subsubsection{Environment and RL task}
\begin{figure}
    \centering
    \includegraphics[width=80mm]{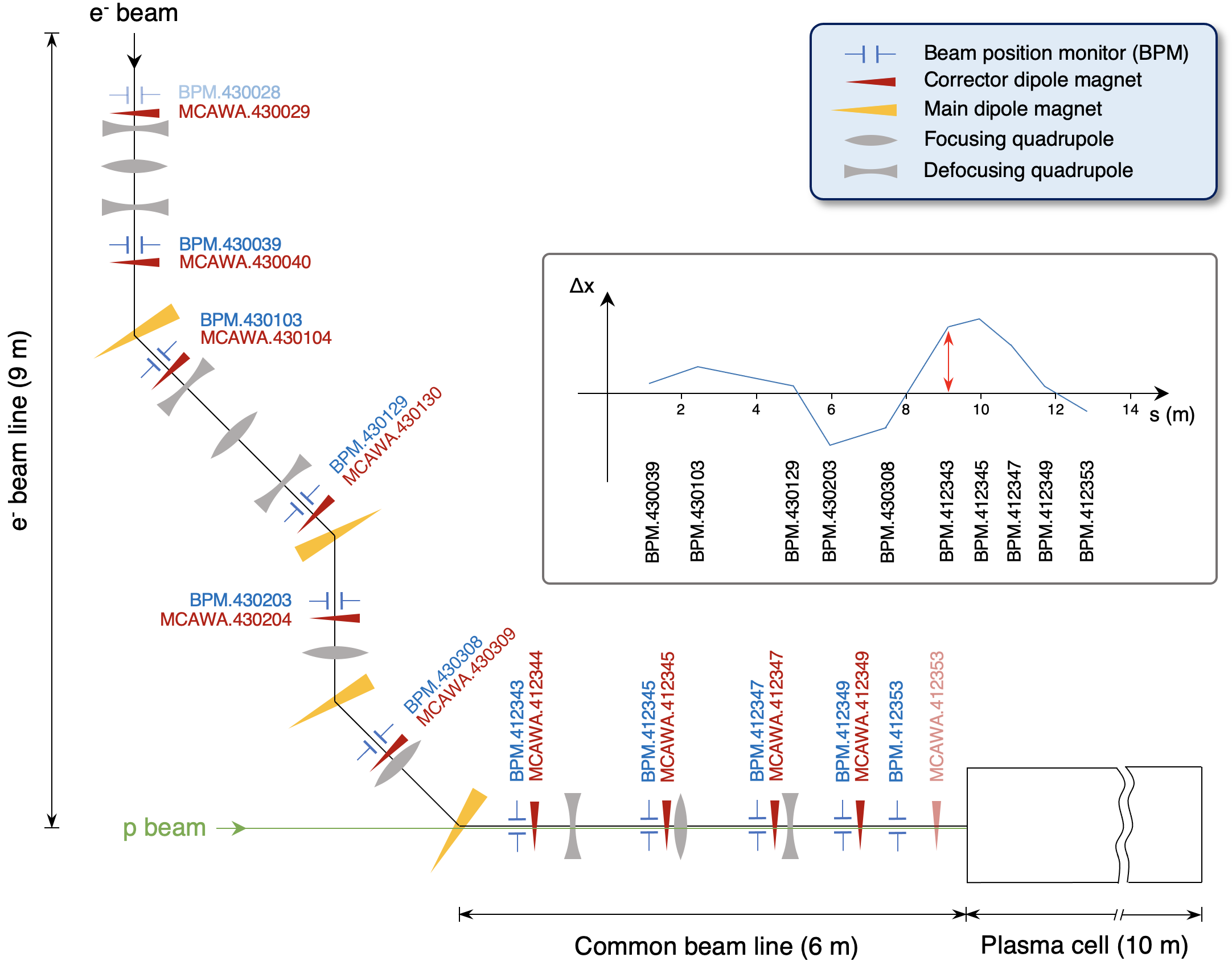}
    \caption{Schematic of the electron beam line of the CERN AWAKE experiment with $10$ trajectory correctors and $10$ beam position readings along the line.}
    \label{fig:10d_env}
\end{figure}

The AWAKE electron source and beam line are particularly interesting for algorithm preparation and testing due to the high repetition rate and insignificant damage potential in case of losing the beam at accelerator components. The AWAKE electrons are generated in a $5\,\text{MV}$ photocathode rf gun, accelerated to $18\,\text{MeV}$ and then transported through a beam line of $12\,\text{m}$ to the AWAKE plasma cell. The trajectory is controlled with $10$ horizontal and $10$ vertical steering dipoles according to the measurements of $10$ BPMs, see Fig.~\ref{fig:10d_env}. The BPM electronic read-out is at $10\,\text{Hz}$ and acquisition through the CERN middleware at $1\,\text{Hz}$. 

A hybrid A-C model was used for trajectory correction on the AWAKE electron beam line with the goal that the trained agents correct the line with similar accuracy as the response matrix-based singular value decomposition (SVD) algorithm that has been traditionally used~\cite{svdOrbit}.

The RL problem of the AWAKE electron beam line is formalised as follows. The state $s$ is defined by a ten-dimensional vector of horizontal beam position differences measured with respect to a reference trajectory. Similarly, the action is a ten-dimensional vector of corrector dipole magnet kick angles within a range of $\pm 300$ \textmu rad. Finally, the reward is defined as the negative root-mean-squared (rms) of the measured beam trajectory with respect to the reference at all the BPMs.

\subsubsection{Results: SQA}
\begin{figure}
    \centering
    \includegraphics[width=85mm]{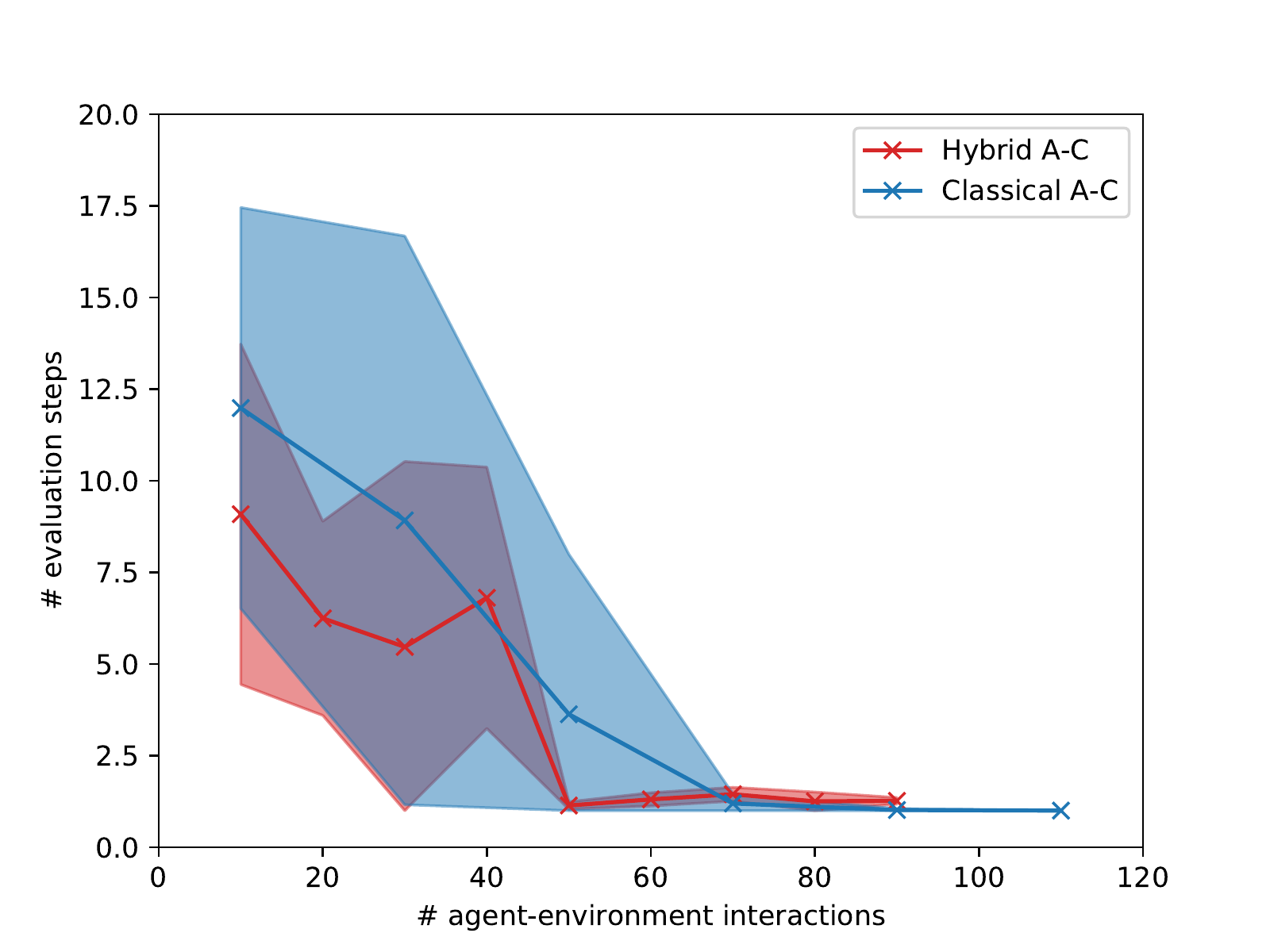}
    \caption{Comparison of the classical (blue) and hybrid (red) A-C algorithms in terms of the number of required agent-environment interactions, i.e.\ sample efficiency, to reach optimal agent behaviour. These are results after extensive hyperparameter tuning. The hybrid A-C has been trained using SQA.}
    \label{fig:actor_critic_sqa_comp}
\end{figure}

\begin{figure*}
    \centering
    \includegraphics[width=145mm]{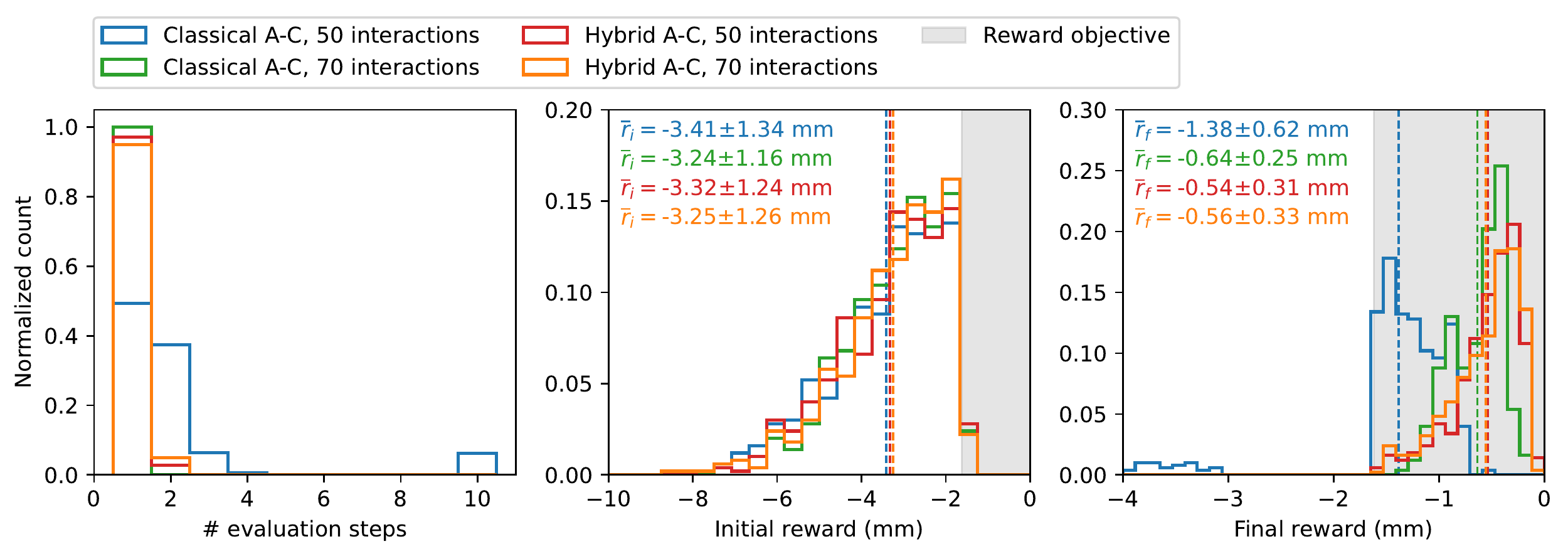}
    \caption{Evaluation of the RL agents trained with a classical A-C algorithm (DDPG) and the hybrid A-C (SQA), respectively, after $50$ and $70$ interactions between agent and environment. Shown are the number of steps taken (left), the distribution of initial rewards (middle), and the distribution of the final rewards at the end of an episode (right). The evaluation has been performed with $500$ randomly initialised episodes.}
    \label{fig:actor_critic_sqa_vs_class_stats}
\end{figure*}

Figure~\ref{fig:actor_critic_sqa_comp} shows the convergence of the classical vs hybrid A-C algorithms after hyperparameter optimisation. $15$ instances of each algorithm were trained and evaluated on $500$ random episodes. The points correspond to the mean number of steps required for the trained actor network to reach the reward objective, set to an equivalent trajectory rms of $1.6\,\text{mm}$, starting from a random initial state. The confidence bound corresponds to the standard deviation over the different instances and variation in the evaluation of episodes. The number of agent-environment interactions does not necessarily correspond to the number of weight updates of the critic and actor networks since a fraction of the agent-environment interactions at the beginning of the training is obtained following a random policy to fill the replay buffer. The fraction of initial random interactions is a hyperparameter tuned with \textsc{Ray Tune}. The QBM consists of $4\times 4$ unit cells of the D-Wave Chimera graph ($8$ qubits each) and features $128$ qubits~\cite{dwavedocs}.

For the case of the AWAKE trajectory steering environment, an improvement in sample efficiency of about $30\,\%$ of the hybrid A-C algorithm over the classical counterpart can be observed. With the hybrid A-C, a near-perfect behaving actor is obtained after training with $50$ agent-environment interactions. The classical algorithm requires about $70$ interactions. The advantage of the hybrid A-C algorithm in terms of sample efficiency may be more apparent for environments with more complex dynamics requiring a lot more agent-environment interactions for successful training. This is currently under study.

Figure~\ref{fig:actor_critic_sqa_vs_class_stats} illustrates the performance differences between the classical and hybrid A-C agents after training for $50$ and $70$ agent-environment interactions, respectively. Histograms show the required number of steps and initial and final rewards during the evaluation phase. Comparing the results for the hybrid A-C after $50$ (red) and $70$ (orange) interactions demonstrates that there is no additional improvement in performance with more training iterations, neither in the number of required steps to reach the objective nor in the final reward values achieved. On the other hand, the classical A-C agent trained for $50$ interactions (blue) has not yet converged. This is visible from the distributions of required steps and the final rewards. In about $8\,\%$ of evaluation episodes, the agent did not manage to solve the task in the $10$ steps available. Accordingly, the final rewards are far from the objective for these cases. Finally, comparing the performance of the classical A-C after $70$ interactions (green) with any of the hybrid A-C results, the classical agent shows a better behaviour in terms of required steps: it always reaches the objective within $1$ step, while the hybrid agents require $2$ steps in about $5\,\%$ to $7\,\%$ of cases. Regarding final rewards, the hybrid agents achieve slightly better mean values than the best classical agent, although with a larger variance. Given the statistical fluctuations, no clear statement can be made here.

\subsubsection{Results: D-Wave Advantage QA} \label{sec:qa_training}
\begin{figure}
    \centering
    \includegraphics[width=80mm]{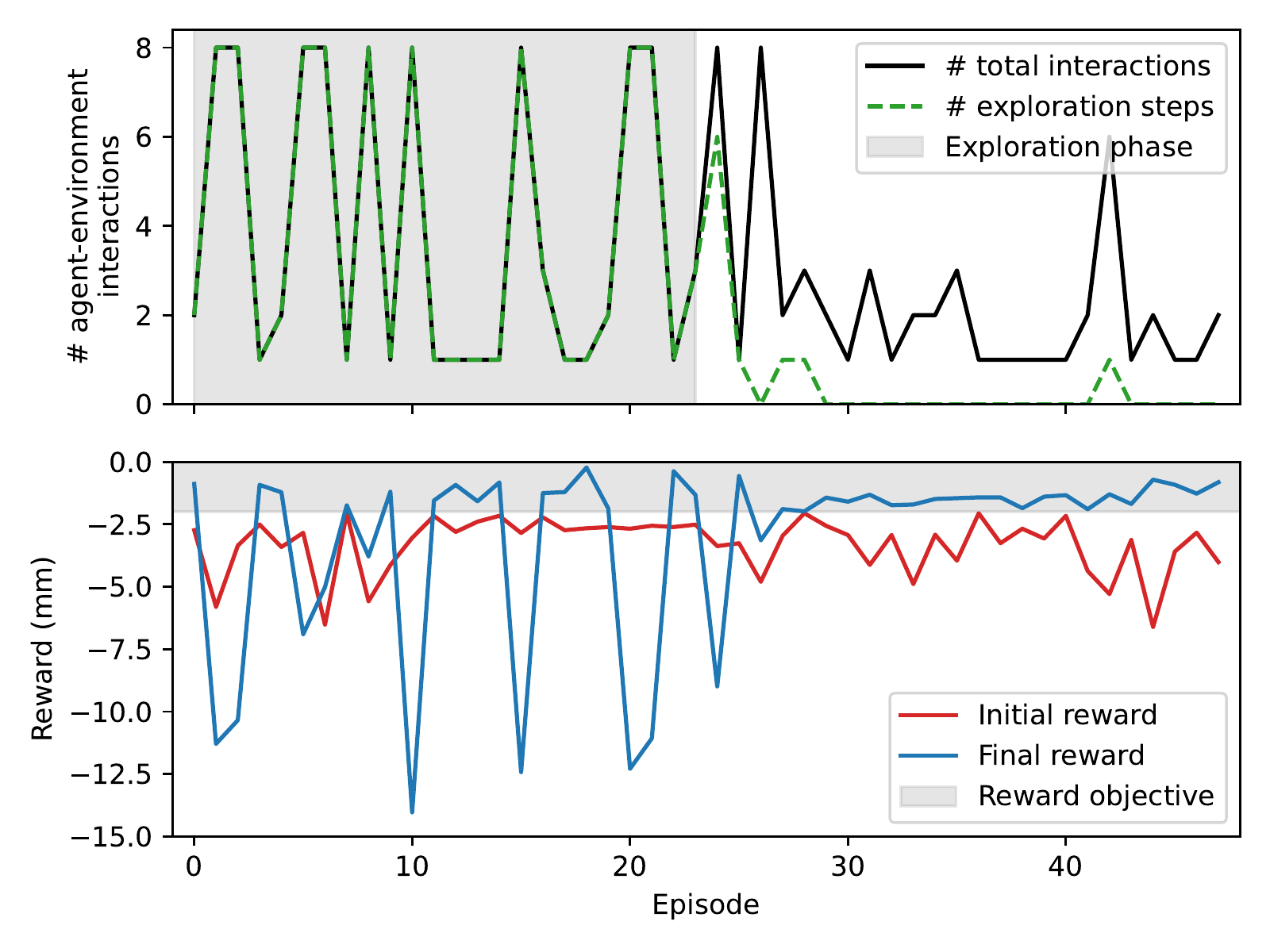}
    \caption{Single RL agent training evolution on D-Wave Advantage Systems using the simulated AWAKE environment with a reward objective of $-2\,\text{mm}$.}
    \label{fig:dwave_awake}
\end{figure}

\begin{figure*}
    \centering
    \includegraphics[width=145mm]{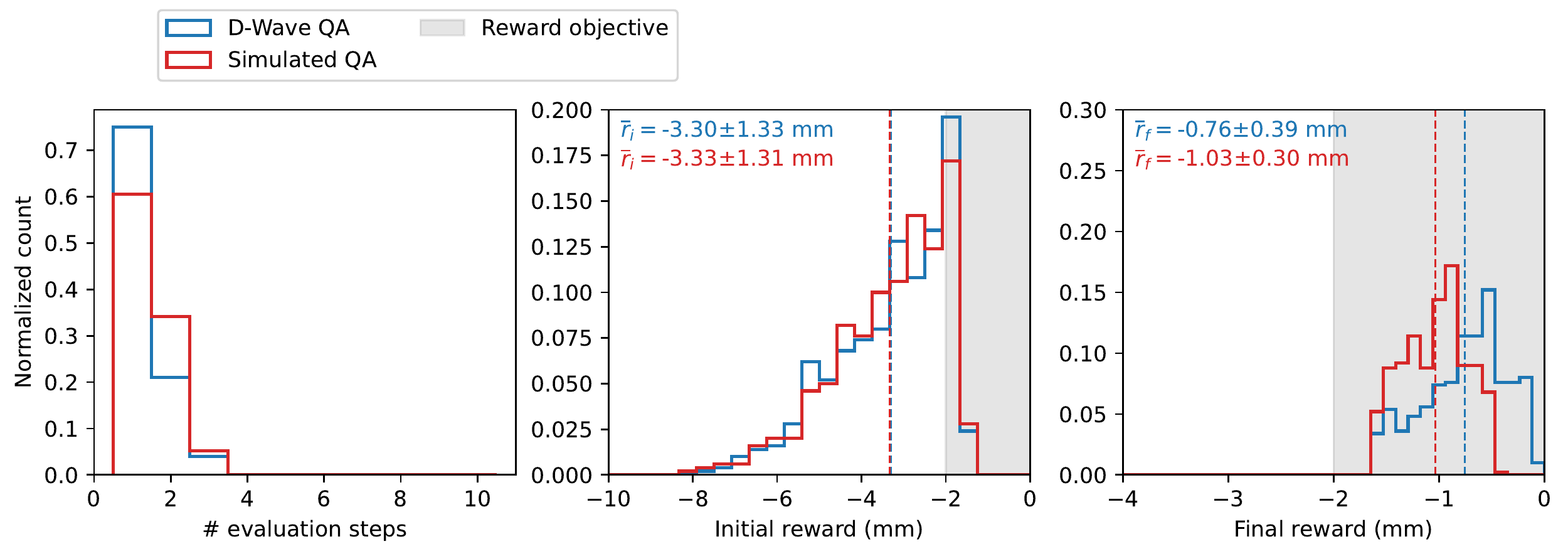}
    \caption{Evaluation of the RL agent trained on D-Wave quantum annealing hardware (blue) and simulated quantum annealing (red) using identical hyperparameters. Shown are the distributions of the number of steps required to reach the reward objective from random initial states (left), as well as the initial (middle) and final (right) rewards.}
    \label{fig:dwave_vs_sqa}
\end{figure*}

The developed hybrid A-C algorithm was also tested on the D-Wave Advantage system. Due to the limited QPU time available, the problem was slightly modified to ensure that it could be solved within the given time frame. To that end, the reward objective was relaxed from an rms value of $1.6\,\text{mm}$ to $2\,\text{mm}$. Using SQA, training parameters, such as the number of agent-environment interactions and replay batch size, were adjusted to adapt the problem to run successfully within the available QPU time.

The training evolution of the RL agent in the simulated AWAKE environment is shown in Fig.~\ref{fig:dwave_awake}. From around episode $30$ onwards, the agent consistently reaches the reward objective. The evaluation of the actor network trained on the D-Wave Advantage (blue) and SQA (red) on $500$ randomly initialised episodes are shown in Fig.~\ref{fig:dwave_vs_sqa}. The plot on the left-hand side displays the number of steps required to reach the reward objective from the initial random state. The middle and right-hand side plots show the distribution of initial and final rewards. The QA-trained agent typically reaches the reward objective within $1$ step ($80\,\%$ of cases). Occasionally, it requires $2$ steps ($18\,\%$), and rarely $3$ steps ($2\,\%$), but it reduces the beam trajectory excursions to an rms better than $2\,\text{mm}$ in $100\,\%$ of cases.

A direct comparison with an RL agent trained with SQA is shown in Fig.~\ref{fig:dwave_vs_sqa}. Both agents were trained with identical hyperparameters. The plots show that their performance is comparable, particularly in terms of the number of steps required to reach the reward objective. The agent trained on the D-Wave QA seems to produce marginally better results on average as can be seen from the distributions of final rewards in Fig.~\ref{fig:dwave_vs_sqa}.

\subsection{Results: deployment on real beam line}
\begin{figure}
    \centering
    \includegraphics[width=80mm]{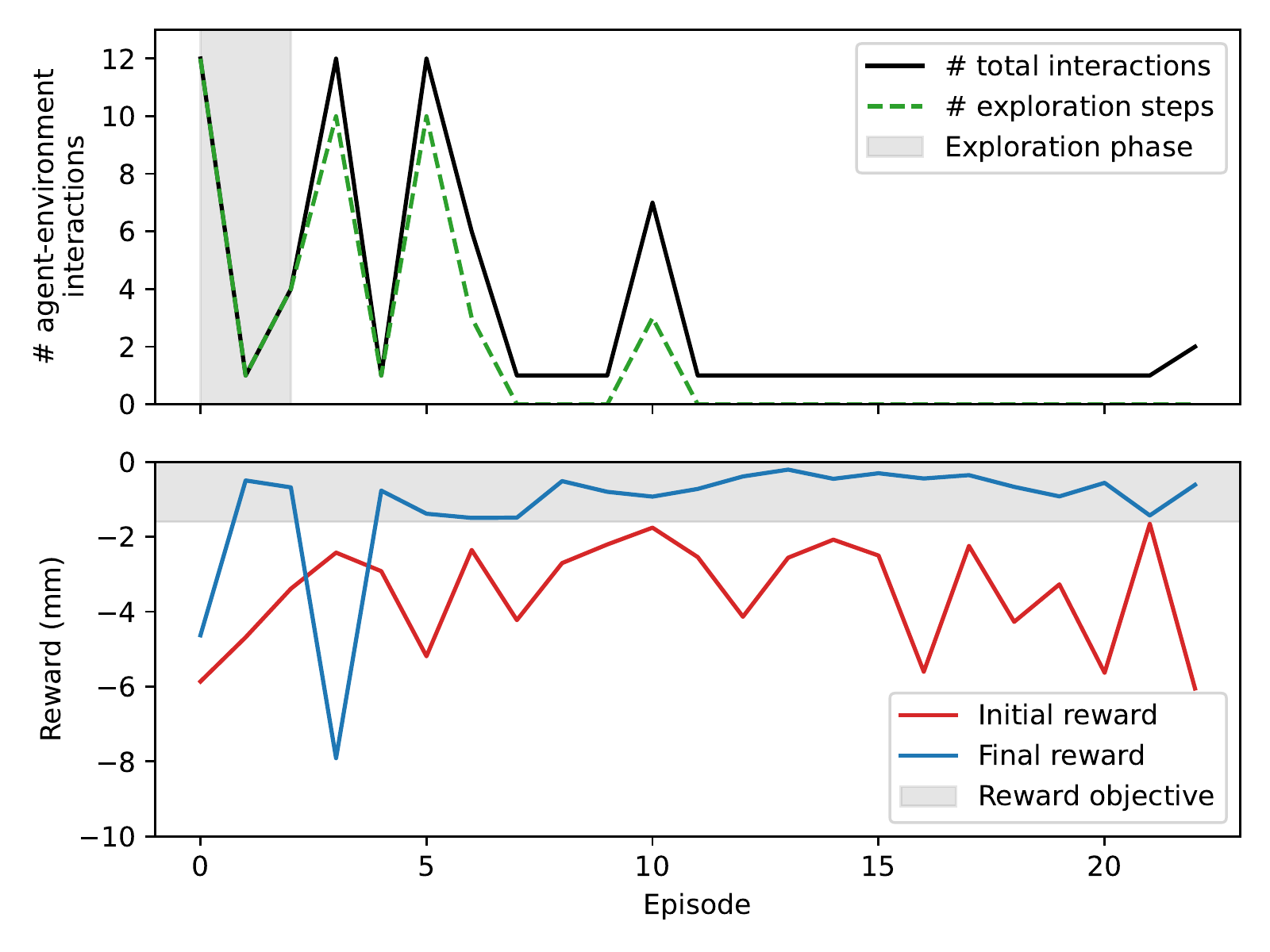}
    \caption{Training of RL agent (SQA) on the simulated AWAKE environment for deployment on the real beam line.}
    \label{fig:real_awake_train}
\end{figure}

\begin{figure*}
    \centering
    \includegraphics[width=145mm]{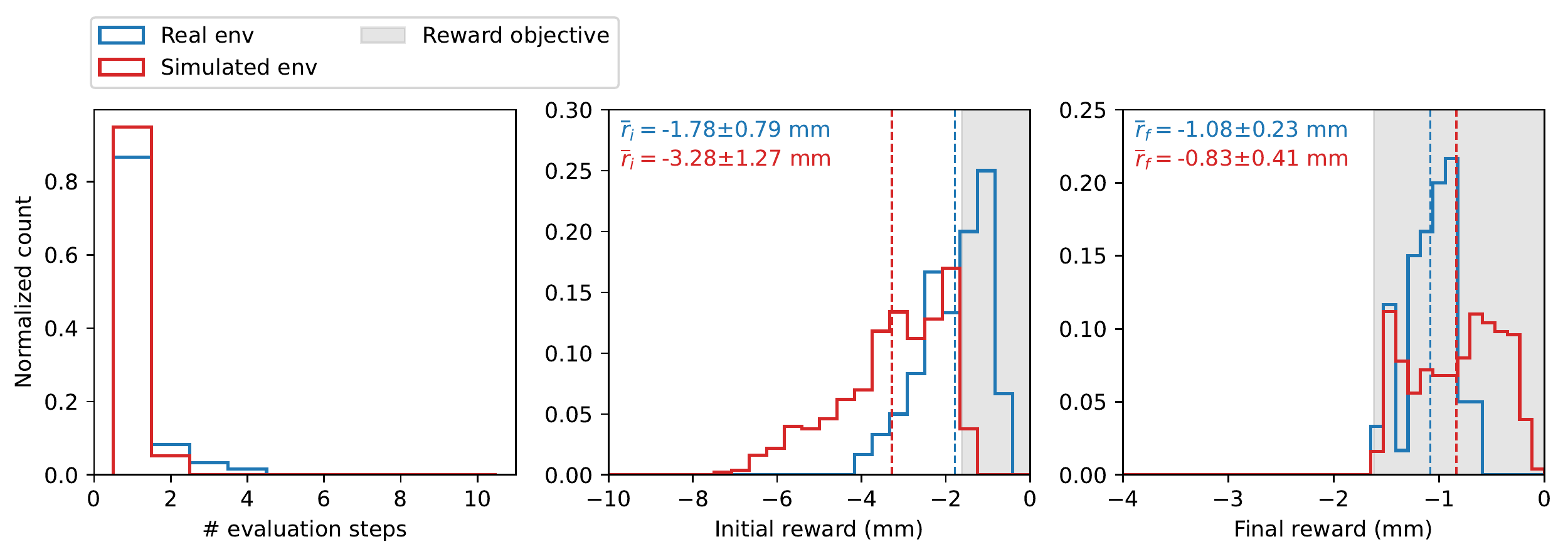}
    \vskip 0.3cm
    \includegraphics[width=145mm]{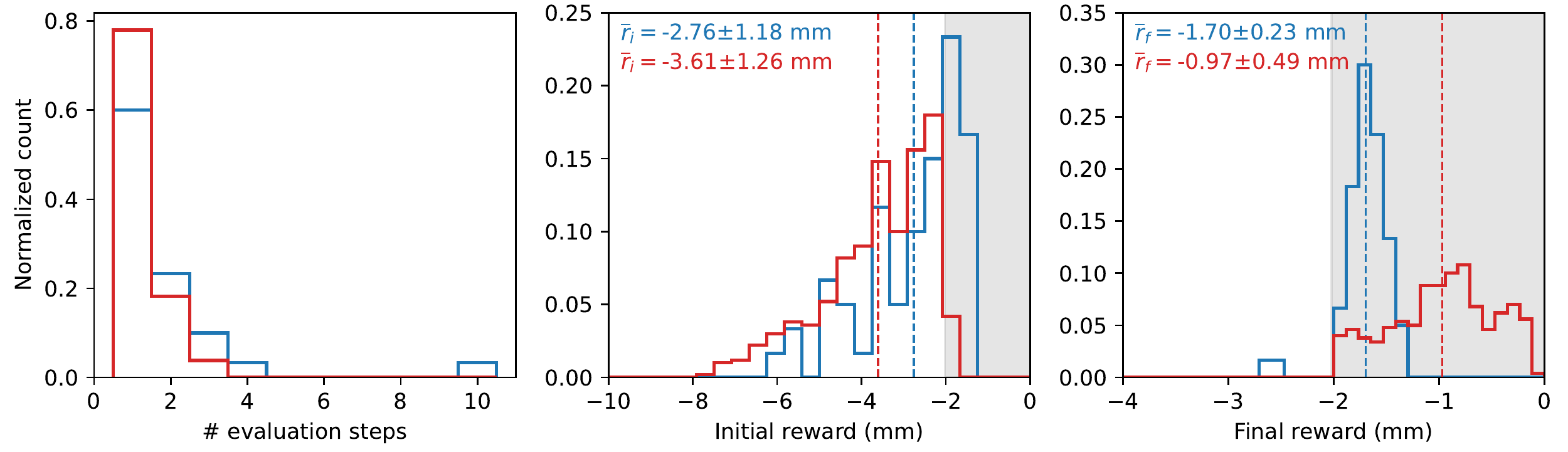}
    \caption{\emph{Top:} evaluation of an SQA-trained agent on the simulated and real AWAKE environments (reward objective: $-1.6\,\text{mm}$). \emph{Bottom:} evaluation of the D-Wave QA-trained agent in the simulated and real AWAKE environments (reward objective: $-2.0\,\text{mm}$). Shown are the distributions of the number of steps required to reach the reward objective from random initial states (left), as well as the initial (middle) and final (right) rewards.}
    \label{fig:real_awake_eval}
\end{figure*}

The hybrid A-C agents were also evaluated in the real AWAKE environment to test sim-to-real transfer. Both agents -- trained on D-Wave QA hardware (Fig.~\ref{fig:real_awake_eval}, bottom row) and with SQA (Fig.~\ref{fig:real_awake_eval}, top row) -- were tested. The training evolution is given in Figs.~\ref{fig:dwave_awake} and~\ref{fig:real_awake_train} with reward objectives set to $-2.0\,\text{mm}$ (QA) and $-1.6\,\text{mm}$ (SQA), respectively. The number of agent-environment interactions is shown (top) together with the evolution of initial and final rewards during the training period. As explained above, using the hybrid A-C scheme, the QBM-based critic is no longer required at inference time since the policy is encoded entirely in the classical actor network.

Figure~\ref{fig:real_awake_eval} shows histograms of the required number of steps as well as initial and final rewards on the simulated (red) and real (blue) AWAKE environments during evaluation. Note that the agents were deployed in the real-world environment without additional training. Both agents can solve the tasks in the simulated and real AWAKE environment. Also, both agents perform better in the simulated environment in terms of final rewards and number of steps required indicating small differences between the simulation and the real beam line. While the measurements on the real beam line were performed back-to-back for the SQA- and QA-trained agents, the distribution of initial rewards (blue, middle column) does not fully overlap, with a tendency towards more challenging episodes for the QA-trained agent (bottom row). The QA-trained agent failed to solve the task for one of the initial states in the real environment, likely indicating that its training had not fully converged due to the limited QPU time available.

\section{Conclusions}\label{sec:conclusions}
Earlier research has demonstrated that free energy-based reinforcement learning (FERL) with quantum Boltzmann machines (QBM) can remarkably increase the sample efficiency compared to the classical deep Q-learning (DQN) algorithm. The results were obtained for control problems defined in discrete action-space environments. This article confirms the improved training efficiency of FERL compared to DQN on the example of the beam trajectory correction task in the TT24-T4 beam line delivering protons to CERN's fixed-target physics experiments. The task was defined using discrete actions but continuous states, extending the applicability of previously reported FERL methods in the literature where the state and action spaces were both discrete. Another interesting result was that FERL learns very efficiently even without experience replay, but the sample efficiency improves further and by a significant amount when including it.

In a second study, this paper investigated whether the increased sample efficiency could be exploited for continuous action-space environments often encountered in real-world control problems. To that end, the authors developed a hybrid RL model that can handle continuous states and actions. Inspired by the actor-critic scheme of the Deep Deterministic Policy Gradient (DDPG) algorithm, the critic network was replaced by a clamped QBM and trained using the FERL approach. Based on the simulated ten-dimensional trajectory correction problem of the AWAKE electron beam line, RL agents were trained with the new method employing simulated quantum annealing and D-Wave Advantage quantum annealing hardware. The agents were then successfully evaluated in the real facility. An increase in sample efficiency was again observed, but the improvement compared to the classical algorithm was not as significant as for the discrete action space problem. Further studies will be carried out to test whether this is generally true or only for the problem studied here, which has linear dynamics. As a final note, the chosen hybrid quantum actor-critic architecture also has a practical advantage -- at inference, i.e.\ in the accelerator control room, only the classical actor network is required which greatly simplifies deployment and usage for accelerator operation.

\bibliographystyle{quantum}
\bibliography{hybrid_ac_quantum_rl_cern}

\begin{thebibliography}{10}

\bibitem{physAtSPS}
L.~Gatignon.
\newblock ``{Physics at the SPS}''.
\newblock \href{https://dx.doi.org/10.1063/1.5016162}{Review of Scientific
  Instruments {\bf 89}, 052501}~(2018).

\bibitem{awake}
E.~Adli, A.~Ahuja, O.~Apsimon, R.~Apsimon, A.-M. Bachmann, et~al.
\newblock ``Acceleration of electrons in the plasma wakefield of a proton
  bunch''.
\newblock \href{https://dx.doi.org/10.1038/s41586-018-0485-4}{Nature {\bf 561},
  363--367}~(2018).

\bibitem{postLIU}
H.~Bartosik and G.~Rumolo.
\newblock ``{Performance of the LHC injector chain after the upgrade and
  potential development}''.
\newblock \url{https://arxiv.org/abs/2203.09202}~(2022).

\bibitem{montbaron}
E.~Montbarbon, J.~Bernhard, D.~Brethoux, M.~Brugger, B.~D. Carlsen, et~al.
\newblock ``{The CERN East Area Renovation}''.
\newblock
  \href{https://dx.doi.org/https://doi.org/10.1016/j.nimb.2019.08.028}{Nuclear
  Instruments and Methods in Physics Research Section B: Beam Interactions with
  Materials and Atoms {\bf 461}, 98--101}~(2019).

\bibitem{suttBart}
R.~S. Sutton and A.~G. Barto.
\newblock ``{R}einforcement {L}earning: {A}n {I}ntroduction''.
\newblock A Bradford Book. Cambridge, MA, USA~(2018).
\newblock Second edition.

\bibitem{mnih}
V.~Mnih, K.~Kavukcuoglu, D.~Silver, A.~Rusu, J.~Veness, et~al.
\newblock ``Human-level control through deep reinforcement learning''.
\newblock \href{https://dx.doi.org/https://doi.org/10.1038/nature14236}{Nature
  {\bf 518}, 529--533}~(2015).

\bibitem{sallHint}
B.~Sallans and G.~E. Hinton.
\newblock ``{R}einforcement {L}earning with {F}actored {S}tates and
  {A}ctions''.
\newblock J. Mach. Learn. Res. {\bf 5}, 1063–1088~(2004).
\newblock
  url:~\href{https://dl.acm.org/doi/10.5555/1005332.1016794}{dl.acm.org/doi/10.5555/1005332.1016794}.

\bibitem{levit}
A.~Levit, D.~Crawford, N.~Ghadermarzy, J.~S. Oberoi, E.~Zahedinejad, and
  P.~Ronagh.
\newblock ``Free energy-based reinforcement learning using a quantum
  processor''.
\newblock \url{https://doi.org/10.48550/arXiv.1706.00074}~(2017).

\bibitem{RL_QBM}
D.~Crawford, A.~Levit, N.~Ghadermarzy, J.~S. Oberoi, and P.~Ronagh.
\newblock ``{R}einforcement {L}earning using quantum {B}oltzmann machines''.
\newblock \href{https://dx.doi.org/10.48550/arXiv.1612.05695}{Quantum Info.
  Comput. {\bf 18}, 51–74}~(2018).

\bibitem{ddpg}
T.~P. Lillicrap, J.~J. Hunt, A.~Pritzel, N.~Heess, T.~Erez, et~al.
\newblock ``Continuous control with deep reinforcement learning''.
\newblock In ICLR (Poster).
\newblock ~(2016).

\bibitem{experienceReplay}
L.-J. Lin.
\newblock ``Self-improving reactive agents based on reinforcement learning,
  planning and teaching''.
\newblock \href{https://dx.doi.org/10.1007/BF00992699}{Mach. Learn. {\bf 8},
  293–321}~(1992).

\bibitem{boltzmannMachine}
G.~E. Hinton and T.~J. Sejnowski.
\newblock ``{O}ptimal {P}erceptual {I}nference''.
\newblock In Proceedings of the IEEE Conference on Computer Vision and Pattern
  Recognition.
\newblock Pages 448--453.
\newblock ~(1983).
\newblock
  url:~\href{https://www.cs.toronto.edu/~hinton/absps/optimal.pdf}{www.cs.toronto.edu/~hinton/absps/optimal.pdf}.

\bibitem{transverseIsingModel}
P.~G. de~Gennes.
\newblock ``Collective motions of hydrogen bonds''.
\newblock \href{https://dx.doi.org/10.1016/0038-1098(63)90212-6}{Solid State
  Communications {\bf 1}, 132--137}~(1963).

\bibitem{QAspins}
M.~Johnson, M.~Amin, S.~Gildert, T.~Lanting, F.~Hamze, et~al.
\newblock ``Quantum annealing with manufactured spins''.
\newblock \href{https://dx.doi.org/10.1038/nature10012}{Nature {\bf 473},
  194--8}~(2011).

\bibitem{farhi2000quantum}
E.~Farhi, J.~Goldstone, S.~Gutmann, and M.~Sipser.
\newblock ``Quantum computation by adiabatic evolution''.
\newblock \url{ https://doi.org/10.48550/arXiv.quant-ph/0001106}~(2000).

\bibitem{born1928beweis}
M.~Born and V.~Fock.
\newblock ``Beweis des adiabatensatzes''.
\newblock \href{https://dx.doi.org/10.1007/BF01343193}{Zeitschrift f{\"u}r
  Physik {\bf 51}, 165--180}~(1928).

\bibitem{cubitt2015undecidability}
T.~S. Cubitt, D.~Perez-Garcia, and M.~Wolf.
\newblock ``Undecidability of the spectral gap''.
\newblock \href{https://dx.doi.org/10.1038/nature16059}{Nature {\bf 528},
  207--211}~(2015).

\bibitem{geoch}
C.~C. McGeoch.
\newblock ``Adiabatic quantum computation and quantum annealing''.
\newblock \href{https://dx.doi.org/10.2200/S00585ED1V01Y201407QMC008}{Synthesis
  Lectures on Quantum Computing}. Morgan \& Claypool Publishers~(2014).

\bibitem{leap}
D-Wave Leap.
\newblock \url{https://cloud.dwavesys.com/leap/}.

\bibitem{biamonte-qml}
J.~Biamonte, P.~Wittek, N.~Pancotti, P.~Rebentrost, N.~Wiebe, and S.~Lloyd.
\newblock ``Quantum machine learning''.
\newblock \href{https://dx.doi.org/10.1038/nature23474}{Nature {\bf 549},
  195--202}~(2017).

\bibitem{PhysRevX.4.031002}
G.~D. Paparo, V.~Dunjko, A.~Makmal, M.~A. Martin-Delgado, and H.~J. Briegel.
\newblock ``{Q}uantum {S}peedup for {A}ctive {L}earning {A}gents''.
\newblock \href{https://dx.doi.org/10.1103/PhysRevX.4.031002}{Phys. Rev. X {\bf
  4}, 031002}~(2014).

\bibitem{dwavedocs}
{D-Wave Systems Inc.}
\newblock ``{D}-{W}ave {S}ystem {D}ocumentation''.
\newblock \url{https://docs.dwavesys.com/}.
\newblock Accessed: 2022-08-04.

\bibitem{jerbi}
S.~Jerbi, C.~Gyurik, S.~Marshall, H.~J. Briegel, and V.~Dunjko.
\newblock ``{P}arametrized {Q}uantum {P}olicies for {R}einforcement
  {L}earning''.
\newblock In Advances in Neural Information Processing Systems 34: Annual
  Conference on Neural Information Processing Systems 2021, NeurIPS 2021,
  December 6-14, 2021, virtual.
\newblock \href{https://dx.doi.org/10.48550/arxiv.2103.05577}{Pages
  28362--28375}.
\newblock ~(2021).

\bibitem{VQCDRL16}
S.~Y.-C. Chen, C.-H.~H. Yang, J.~Qi, P.-Y. Chen, X.~Ma, and H.-S. Goan.
\newblock ``{V}ariational {Q}uantum {C}ircuits for {D}eep {R}einforcement
  {L}earning''.
\newblock \href{https://dx.doi.org/10.1109/ACCESS.2020.3010470}{IEEE Access
  {\bf 8}, 141007--141024}~(2020).

\bibitem{RLQVC17}
O.~Lockwood and M.~Si.
\newblock ``{R}einforcement {L}earning with {Q}uantum {V}ariational
  {C}ircuits''.
\newblock In Proceedings of the Sixteenth AAAI Conference on Artificial
  Intelligence and Interactive Digital Entertainment.
\newblock \href{https://dx.doi.org/10.48550/arXiv.2008.07524}{AIIDE'20}. AAAI
  Press~(2020).

\bibitem{Jerbi_2021}
S.~Jerbi, L.~M. Trenkwalder, H.~Poulsen~Nautrup, H.~J. Briegel, and V.~Dunjko.
\newblock ``{Q}uantum {E}nhancements for {D}eep {R}einforcement {L}earning in
  {L}arge {S}paces''.
\newblock \href{https://dx.doi.org/10.1103/PRXQuantum.2.010328}{PRX Quantum
  {\bf 2}, 010328}~(2021).

\bibitem{dwave66}
F.~Neukart, D.~Von~Dollen, C.~Seidel, and G.~Compostella.
\newblock ``{Q}uantum-{E}nhanced {R}einforcement {L}earning for
  {F}inite-{E}pisode {G}ames with {D}iscrete {S}tate {S}paces''.
\newblock \href{https://dx.doi.org/10.3389/fphy.2017.00071}{Frontiers in
  Physics{\bf 5}}~(2018).

\bibitem{openai}
G.~Brockman, V.~Cheung, L.~Pettersson, J.~Schneider, J.~Schulman, J.~Tang, and
  W.~Zaremba.
\newblock ``Openai gym''~(2016).
\newblock
  url:~\href{https://doi.org/10.48550/arXiv.1606.01540}{doi.org/10.48550/arXiv.1606.01540}.

\bibitem{stable-baselines3}
A.~Raffin, A.~Hill, A.~Gleave, A.~Kanervisto, M.~Ernestus, and N.~Dormann.
\newblock ``{S}table-{B}aselines3: {R}eliable {R}einforcement {L}earning
  {I}mplementations''.
\newblock Journal of Machine Learning Research {\bf 22}, 1--8~(2021).
\newblock
  url:~\href{http://jmlr.org/papers/v22/20-1364.html}{http://jmlr.org/papers/v22/20-1364.html}.

\bibitem{sqaod}
S.~Morino.
\newblock ``Sqaod: simulated quantum annealing library''.
\newblock \url{https://github.com/shinmorino/sqaod}.
\newblock Accessed: 2022-08-04.

\bibitem{raytune}
R.~Liaw, E.~Liang, R.~Nishihara, P.~Moritz, J.~E. Gonzalez, and I.~Stoica.
\newblock ``{T}une: {A} {R}esearch {P}latform for {D}istributed {M}odel
  {S}election and {T}raining''.
\newblock \url{https://doi.org/10.48550/arXiv.1807.05118}~(2018).

\bibitem{optuna}
T.~Akiba, S.~Sano, T.~Yanase, T.~Ohta, and M.~Koyama.
\newblock ``{O}ptuna: {A} {N}ext-generation {H}yperparameter {O}ptimization
  {F}ramework''.
\newblock \href{https://dx.doi.org/10.48550/arXiv.1907.10902}{CoRR}~(2019).

\bibitem{tt24t4}
G.~L. D'Alessandro, D.~Banerjee, J.~Bernhard, M.~Brugger, N.~Doble, et~al.
\newblock ``{Target Bypass Beam Optics for Future High Intensity Fixed Target
  Experiments in the CERN North Area}''.
\newblock In Proc. IPAC'21.
\newblock \href{https://dx.doi.org/10.18429/JACoW-IPAC2021-WEPAB185}{Pages
  3046--3048}.
\newblock Number~12 in International Particle Accelerator Conference. JACoW
  Publishing, Geneva, Switzerland~(2021).

\bibitem{xband}
R.~Agustsson, R.~Carriere, O.~Chimalpopoca, V.~Dolgashev, M.~A. Gusarova, S.~V.
  Kutsaev, and A.~Yu~Smirnov.
\newblock ``{Experimental studies of a high-gradient X-band welded hard-copper
  split accelerating structure}''.
\newblock \href{https://dx.doi.org/10.1088/1361-6463/ac4632}{Journal of Physics
  D: Applied Physics {\bf 55}, 145001}~(2022).

\bibitem{svdOrbit}
Y.~Chung, G.~Decker, and K.~Evans.
\newblock ``Closed orbit correction using singular value decomposition of the
  response matrix''.
\newblock \url{https://www.osti.gov/servlets/purl/10178130}~(1993).
\newblock Argonne National Laboratory. Unpublished.

\end{thebibliography}

\onecolumn
\appendix
\section{Complementary results} \label{app:1ddiscretebinary}
\subsection{FERL Q-learning with discrete, binary state-action space}
\begin{figure}[b]
    \centering
    \includegraphics[width=65mm]{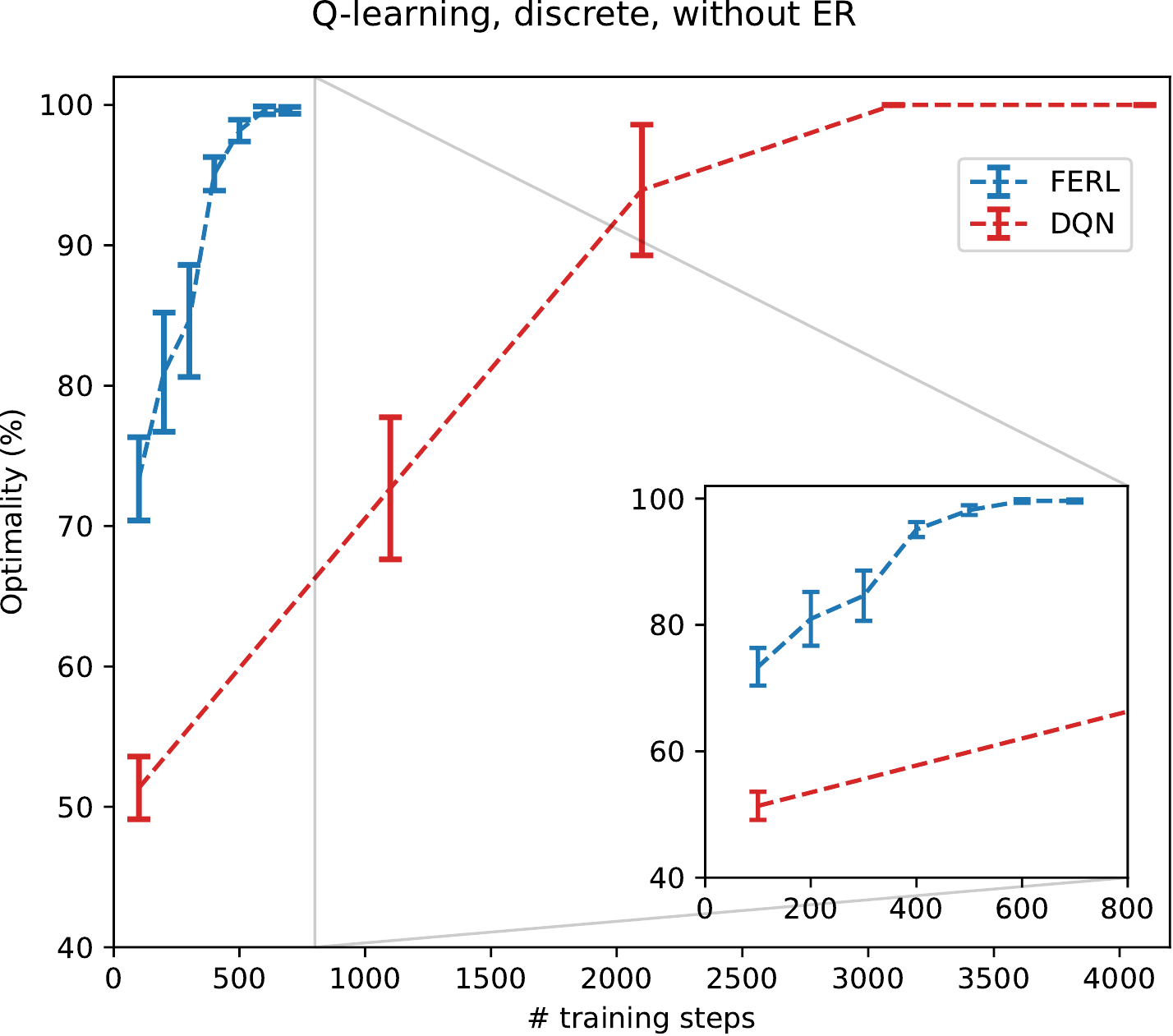}
    \hspace{0.8cm}
    \includegraphics[width=65mm]{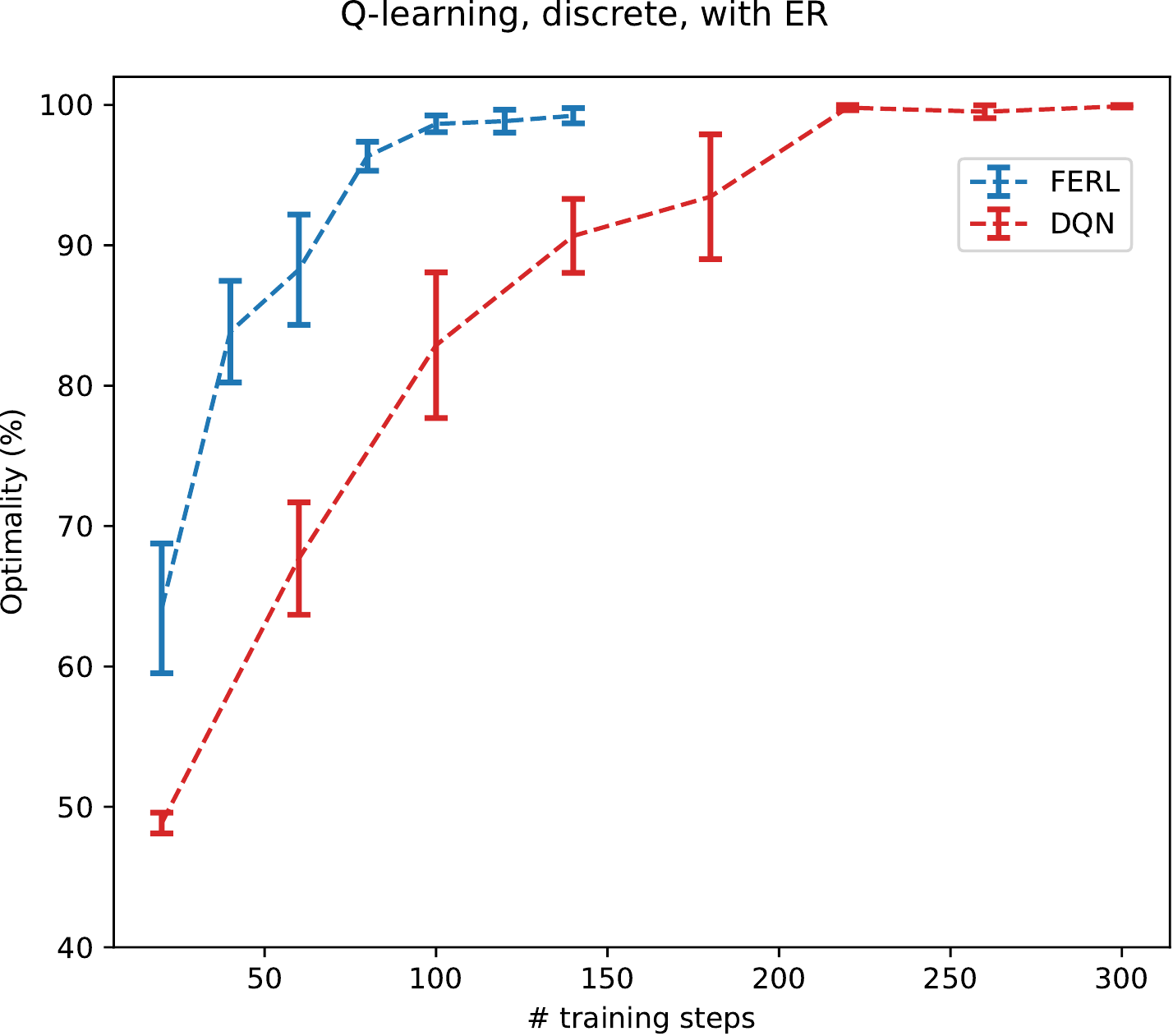}
    \caption{One-dimensional beam steering environment with a discrete, binary state-action space: convergence study for agent optimality vs the number of training steps \emph{without} (left) and \emph{with} (right) experience replay for a classical DQN (red) and an FERL (blue) agent shown on a log-log scale.}
    \label{fig:1d_discr}
\end{figure}

Figure~\ref{fig:1d_discr} shows the results of policy optimality vs number of training steps with the classical deep Q-learning (red) and FERL (blue) agents, without (left) and with (right) experience replay. Again, the study is based on the one-dimensional TT24-T4 target beam steering environment described in Section~\ref{sec:1dsteering}. However, this time it uses a discrete, binary state-action space, in analogy to the studies in~\cite{levit}.

Similarly to Fig.~\ref{fig:1d_cont}, which was obtained for a continuous state space, the results here show that the FERL agent manages to outperform the classical deep Q-learning agent. However, the gain in sample efficiency is much smaller than in the former case. Again, the comparison between the left and right plots confirms the importance of experience replay for both types of algorithms. In the best case, the FERL algorithm reliably reaches $100\,\%$ optimality with about $100$ training steps, while classical deep Q-learning requires about $220$ steps.

\end{document}